\documentclass[aps,prd,nofootinbib,showpacs,preprintnumbers,longbibliography,amssymb,floatfix,11pt]{revtex4-1} % not twocolumn

\usepackage[sort&compress]{natbib}
\usepackage[dvipsnames]{xcolor}

\usepackage{graphicx}% Include figure files
\usepackage{epsfig}
\usepackage{dcolumn}% Align table columns on decimal point
\usepackage{bm}% bold math

\usepackage{amsmath}
\usepackage{siunitx}
\usepackage{amssymb}
\usepackage{physics}
\usepackage{slashed}
\usepackage{dsfont}
\usepackage{subcaption}
\usepackage{hyperref}
\usepackage{cleveref}
\usepackage{soul}

\hypersetup{
    colorlinks=true,       % false: boxed links; true: colored links
    linkcolor=purple,   %black,          % color of internal links
    citecolor=blue,        % color of links to bibliography
    filecolor=magenta,      % color of file links
    urlcolor=blue           % color of external links
}
\graphicspath{{Figures/}}

\begin{document}

	\preprint{MITP-24-064}

	\title{Interference Resurrection of the $\tau$ Dipole through Quantum Tomography}

	\author{Prisco Lo Chiatto\footnote{Now at 
			    Max-Planck-Institut für Physik (MPP),
            Föhringer Ring 6, 80805 Munich, Germany
			and
    Weizmann Institute of Science, P.O. Box 26, 76100 Rehovot}}
	\email{prisco.lochiatto@mpp.mpg.de}
	\affiliation{PRISMA$^+$ Cluster of Excellence \& Mainz Institute for Theoretical Physics\\
	Johannes Gutenberg University, 55099 Mainz, Germany}

	\begin{abstract}
		Helicity selection rules can suppress the leading contributions from dimension-6 operators in the Standard Model Effective Field Theory (SMEFT), reducing sensitivity to potential new physics. This paper explores how the interference contributions from the $\tau$ lepton's anomalous dipole moment is restored in different observables, in particular comparing the sensitivity to SMEFT operators of quantum information observable and more traditional spin correlations. We compute the sensitivity of various observables—including entanglement measures, Bell inequality violations, and quantum uncertainties—to new physics effects using Monte Carlo simulations. Spin correlation observables are found to outperform both the integrated cross-section and quantum information observables in sensitivity to both $CP$-conserving and -violating effects, improving the sensitivity to the scale of new physics by up to a factor of 3. Our results suggest that quantum information observables are suboptimal probes for resurrecting the interference, and more in general to disentangle the $CP$ properties of new physics.
\end{abstract}

	\maketitle
	%%%%%%%%%%%%%%%%%%%%%%%%%%%%%
	\section{Introduction}
	\label{sec:Intro}
	%%%%%%%%%%%%%%%%%%%%%%%%%%%%%
	The search for deviations from the Standard Model (SM) has entered an era of precision tests. New physics (NP) whose characteristic scale $\Lambda$ is well-separated from the electroweak (EW) scale $v$ can be described with the tools of effective field theory (EFT) to a largely model-independent extent. The Standard Model effective field theory (SMEFT)~\cite{Buchmuller:1985jz,Brivio:2017vri} is one such EFT, which encodes the effect of heavy NP in a tower of operators, organised according to the number of derivatives and fields. In most cases, naive dimensional analysis (NDA) can be used to argue that the leading deviation from the SM prediction for a given observable is due to operators with canonical mass dimension 6. In particular, the leading deviation from the SM in observables is expected to arise at order $1/\Lambda^2$, corresponding to the interference between the SM matrix at dimension-6. However, (approximate) symmetries and selection rules can cause this expectation to fail~\cite{Falkowski:2016cxu,Liu:2016idz,Azatov:2016sqh}. In such cases, operators with higher mass dimension, loop corrections and multiple insertions of dimension 6 operators can have parametrically larger contributions than the interference. In Ref.~\cite{Azatov:2016sqh}, it was shown that a selection rule arises whenever the SM and SMEFT contribute to mutually exclusive helicity amplitudes. In the high-energy limit, the chiral nature of the SM then forces certain helicity amplitudes to vanish, so selection rules that are only lifted by finite masses emerge.

	In cases where noninterference happens, the sensitivity that a given observable has to NP effects is lowered by additional inverse powers of the large scale $\Lambda$. Noninterference can be circumvented by looking at different observables or different final states, in a process dubbed ``interference resurrection.'' For instance, if the final states are allowed to decay, the full process does not need to respect the helicity selection rule, and the interference is resurrected. One phenomenologically relevant case that has received much attention is triple gauge coupling, both for the triple-W~\cite{Helset:2017mlf,Azatov:2019xxn,ElFaham:2024uop} and triple-gluon~\cite{Dixon:1993xd} case. 

	In this paper, we consider resurrection in 4-fermion scattering, which is relevant for searches at both lepton and hadron collider experiments. 
	We show, through a quantum mechanical example, that the measurement of the spin correlations of final-state leptons in the process $f\bar f\to \ell \bar \ell$ can resurrect the interference of operators that generate electric and magnetic dipole moments, elucidating at the same time why the same procedure fails with other noninterfering operators. We therefore focus on operators contributing to the anomalous dipole moment of the $\tau$, $g_{\tau}-2$. Unlike the anomalous dipole moment of the lighter leptons, the short lifetime of the $\tau$ prevents $g_{\tau}-2$ from being measured at low energies. Viable probes are then forced to be at high energies, and as such they suffer from noninterfence. Spin correlation then offer an interesting window on the anomalous dipole moments by resurrecting the interference. 

	Recently, the study of spin correlations at colliders has gained renewed interest, after the prediction~\cite{Afik:2020onf} and successful measurement~\cite{ATLAS:2023fsd} of entanglement between the spins of a top quark pair at the Large Hadron Collider (LHC), as discussed in the review paper Ref.~\cite{Barr:2024djo}. Following the formalism of the celebrated Bell inequalities~\cite{Bell:1964kc}, and of the groundbreaking experimental and theoretical efforts of the 1960s~\cite{PhysRevLett.28.938,CHSH_1969}, experiments known as ``Bell tests'' were developed. They aim to exclude hidden variable theories by measuring correlations between spins that could not arise in theories based on local realism.
	At colliders, the correlations between the polarisations of the final states from hard scattering events are measured only indirectly. For unstable particles, the angular distribution decay product can be used instead, relying on angular momentum conservation.\footnote{This measurement strategy is known to generate a loophole in Bell tests, see Refs.~\cite{Abel:1992kz,Li:2024luk}. Since the focus of this paper is not a test of quantum mechanics, this will not be an issue for us.} This also explains, in part, why the measurement of spin correlations can resurrect the interference.  

	From a phenomenological and model building point of view, much of the recent effort on observables related to entanglement has been on establishing their sensitivity to physics beyond the Standard Model (BSM). For top quark pair production, it has been suggested~\cite{Maltoni:2024tul} that quantum information observables either surpass or complement the sensitivity of both the integrated cross-section and other angular observables.
This seems to tie in perfectly with the fact that spin correlations can be used for interference resurrection, enhancing the sensitivity with respect to the cross-section.
However, as we will elaborate in the following, in the case of top pair production the absence of $CP$ violation allows for a particularly simple combination of spin correlation coefficients to be used as a marker for entanglement.
This variable had already been introduced as a good tool to trace spin correlations in Ref.~\cite{Bernreuther:2004jv} and used as a probe for NP in Ref.~\cite{Bernreuther:2017yhg} without any need to reference quantum information.
To test whether access to uniquely quantum phenomena is a necessary ingredient for increased sensitivity, we then evaluate the reach of different quantum information observables, and compare it to the spin correlation coefficients, which when taken individually do not allow to test whether the quantum state has non-classical properties. 
We show that quantum information observables are suboptimal probes for 4-fermion operators, when compared with even a single spin correlation coefficient, and that they partly erase the CP information. 
Indeed, different choices for the phase of the Wilson coefficients lead to polarisation along different axes. 
Quantum information observables, combining all spin correlation in one measure, are unable to differentiate between different choices for the phase of the Wilson coefficients, causing the information about $CP$ properties to be lost. 
We also evaluate the possible reach of present and future colliders to the dipole operators of the $\tau$.

The paper is structured as follows: in \cref{sec:resurrection}, we introduce noninterference and resurrection of interference, then use a simple quantum mechanical example to explain the features of the resurrection mechanism. 
In \cref{sec:formalism} we introduce the formalism to describe spin correlations, as well as the quantum information observables. 
In \cref{sec:results} we analytically calculate the spin observables for both lepton and hadron colliders at tree level and without including $\tau$ decays, and validate the results for lepton colliders against  Monte Carlo simulations that include decays, as well as detector and initial state radiation (ISR) effects. We also critically analyse the NP reach of spin observables. We conclude in \cref{sec:conclusions}. The Fano coefficients of the $\tau\tau$ density matrix are collected in \cref{sec:Fano}. \cref{sec:DetailsMC} and \cref{sec:ImpactPar} respectively contain the results of the MC simulations and the procedure to reconstruct the $\tau$ rest frame using decays with one unobserved $\nu$.

	\section{Interference Suppression and Resurrection in 4-Fermion Scattering}
	\label{sec:resurrection}
	Heavy BSM physics generically impacts low energy observables through virtual corrections, which in the infrared are indistinguishable from higher dimensional operators; this observation, coupled with the experimental fact that the symmetries of the SM seem to be a valid at low energy, lets us organise BSM contributions in a largely model-independent way by considering all operators in an EFT built up from the SM fields in increasing mass dimension~\cite{Buchmuller:1985jz,Brivio:2017vri}. We will adopt the SMEFT in this paper to discuss new physics effects, under the assumption that the underlying theory lives at a scale $\Lambda \gg v$. 

	In this paper, we will be interested in $f\bar f \to \ell \bar \ell$, with $f$ a charged SM fermion and $\ell$ a charged SM lepton. At dimension 6 --- the lowest dimension higher than 4 to affect this process at LO --- the full list of SMEFT operator is well-known~\cite{Buchmuller:1985jz,Grzadkowski:2010es}. Truncating the operator basis at dimension 6, we can compute the squared matrix element. It contains both SM and SMEFT contributions:
	\begin{align}
		\mathcal{M} &= \mathcal{M}^{\textrm SM} + \frac{1}{\Lambda^2}\mathcal{M}^{\textrm SMEFT} \, ,\\
		|\mathcal{M}|^2 &= |\mathcal{M}^{\textrm SM}|^2 + 2 \Re\left(  \frac{1}{\Lambda^2}\mathcal{M}^{\textrm SM}{\mathcal{M}^{\textrm SMEFT}}^* \right) + \frac{1}{\Lambda^4}|\mathcal{M}^{\textrm SMEFT}|^2 \, , \\
				&\equiv {|\mathcal{M}|^2}^{(0)} + \frac{1}{\Lambda^2}{|\mathcal{M}|^2}^{(2)} + \frac{1}{\Lambda^4}{|\mathcal{M}|^2}^{(4)} \, .
				\label{eq:MatEl}
	\end{align}
	Using NDA, one can estimate the energy scaling for ${|\mathcal{M}|^2}^{(n)}$; the interference contribution gives in most cases the leading deviation from the SM, simply because it carries fewer powers of the small ratio $(E/\Lambda)$ than the SMEFT squared piece, where $E$ is the energy scale of the hard interaction. However, as will be discussed shortly, selection rules can make the interference contribution vanish, rendering the SMEFT squared piece the leading deviation from the SM\@. In this case, the experimental sensitivity to this operator is expected to be much weaker than if the selection rule were absent, due to the energy suppression.\footnote{Note that, if the dimension-6 squared contributions are considered, then the contributions of dimension-8 and loop-induced contributions can be of the same order and should generally to be considered when evaluating the sensitivity. This is beyond the scope of this paper.} Such a situation has been known since the 1990s to arise in the case of triple-gluon interaction~\cite{Simmons:1989zs,Simmons:1990dh}, and strategies to circumvent the suppression where quickly proposed~\cite{Dixon:1993xd}. Recently, this topic has attracted renewed interest~\cite{Azatov:2016sqh,Azatov:2017kzw,Falkowski:2016cxu,Aoude:2019cmc}, mostly for the phenomenologically important case of pair production of electroweak bosons; in this context, strategies to circumvent the helicity suppression go under the name of ``interference resurrection''~\cite{Panico:2017frx}, that we will use throughout this paper.

	Because the SM is a chiral theory, 4-fermion effective operators exhibit accidental selection rules in the limit of vanishing fermion masses, where helicity coincides with chirality. For instance, the three-point operators $f\bar fV$ ($V= \gamma, Z$), only allow $f$ and $\bar f$ to have opposite helicities, if the vector is transversely polarised. That is, the helicity matrix element $\mathcal{M}_{h_V h_f,h_{\bar f}}$, with $h_{X}$ the helicity of particle $X$, satisfies
	\begin{equation}
		\mathcal{M}_{T - -} = \mathcal{M}_{T + +} = 0\, .
		\label{eq:Hel3PtSM}
	\end{equation}
	This implies specific helicity configurations for SM lepton pair production $f \bar f \to \ell \bar \ell$ mediated by electroweak gauge bosons at tree level. We will be interested in the cases $f = e,\mu,u,d,s$, all of which have small masses; since the coupling of fermions to longitudinally polarised gauge boson scales with the mass, we only need to consider transverse polarisation of the gauge boson. Then, the helicity matrix element $M_{h_f h_{\bar f} h_{\ell} h_{\bar \ell}}$ is nonvanishing only when $h_f(h_\ell)$ is opposite to $h_{\bar f}(h_{\bar \ell})$. In the following, we will denote this by saying that the SM imposes the helicity structure $(h_f,h_{\bar f},h_\ell,h_{\bar \ell} ) =  (+ - + -), (- + + -), (+ - - +),(- + - +)$.  For nonzero lepton masses, the other helicity structures are suppressed by powers of $m_\ell/m_{\ell\ell}$,  where $m_{\ell \ell}$ is the invariant mass of the di-lepton pair, and the longitudinal polarisation of the gauge bosons are likewise suppressed. In conclusion, in the SM:
	\begin{equation}
		\mathcal{M}_{h_f h_{\bar f} h_{\ell} h_{\bar \ell}}^{\textrm SM} = 0 + \mathcal{O}(\frac{m_\ell}{m_{\ell \ell}}) \quad \textrm{if}\; h_{\ell} = h_{\bar \ell}\; \textrm{or}\, h_{f} = h_{\bar f}\, .
		\label{eq:Hel4ptSM}
	\end{equation}

	Given that SM leptons are light, at energies $m_{\ell \ell} \gtrsim 40$ GeV helicity-violating contributions are heavily suppressed even for the heaviest lepton, the $\tau$. We refer to this as helicity suppression, and we call helicity selection rule the emerging selection rule in the exact chiral limit.
	\begin{table}
		\centering
		$\begin{array}{|c|c|c|}
			\hline
			\text{Operator} & \text{Helicity Structure}\\
			\hline
			\bar \ell \gamma_\mu \ell A^\mu, \bar \ell \gamma_\mu \ell Z^\mu, \bar \ell \gamma_\mu\gamma^5 \ell A^\mu,\bar \ell \gamma_\mu\gamma^5 \ell Z^\mu& (+ - T), (- + T)\\
			\mathcal{O}_{\ell\ell}, \mathcal{O}_{\ell edq},\mathcal{O}_{\ell u}, \mathcal{O}_{\ell q}^{1}, \mathcal{O}_{\ell q}^{3} & (+ - + -), (- + + -), (+ - - +),(- + - +)\\
			\mathcal{O}_{\ell equ}^{1}, \mathcal{O}_{\ell equ}^{3}& (+ + + +), (- - - -)\\
			\mathcal{O}_{W\ell}, \mathcal{O}_{B\ell}& (+ + T), (- - T)\\
			\hline
		\end{array}$
		\caption{Operators that enter $\bar f f \to \bar \ell \ell$ up to dimension 6 in the SMEFT, and their helicity structure in the chiral limit. For the SMEFT operators we follow the notation of \cite{Grzadkowski:2010es}.}
		\label{tab:Hels}
	\end{table}

	We now add dimension-6 SMEFT operators, whose helicity structures are summarised in \cref{tab:Hels}. Most operators that enter lepton pair production at dimension 6 in the SMEFT are unconstrained by helicity selection rules, because they have the same helicity structure as the SM\@.  Only two classes of operators suffer from helicity suppression in the interference: the left-right quark-lepton 4-fermion operators $\mathcal{O}_{\ell equ}^{1},\mathcal{O}_{\ell equ}^{3}$, and the dipole operators $\mathcal{O}_{W\ell},\mathcal{O}_{B\ell}$. In both cases, in the chiral limit it is impossible to construct a non-zero interference $\mathcal{M}_{h_{\ell}h_{\bar \ell}}^{(\textrm SM)}{\mathcal{M}_{h_{\ell}h_{\bar \ell}}^{(2)}}^*$. We underline that in the calculation of the (differential) cross-section for on-shell particles, no product of matrix elements with different helicities can appear, but as we will see this is not true for a generic observable. Indeed, the only way to resurrect interference is to find an observable that is built out of different helicity structures. In the next subsection, using a simple quantum mechanical example, we argue that spin observables can indeed display resurrection of interference.

	\subsection{Resurrection in Spin Observables: a Quantum Mechanics Example}
	\label{sec:ToyModel}
	Since leptons are spin 1/2 particles, the spin state of a $\ell \bar \ell$ pair is represented by a bipartite two-qubit. That is, if we consider the initial state $\ket{f \bar f}$ to be a pure state, the final state can be expressed as:
	\begin{equation}
		\ket{\psi}= S\ket{f\bar f} = \sum_{s_\ell,s_{\bar \ell}} \mathcal{M}_{s_{\ell}s_{\bar \ell}} \ket{s_{\ell}s_{\bar \ell}} + \ldots \, ,
		\label{eq:psi}
	\end{equation}
	with $S$ the scattering matrix, and the dots stand for all possible final states other than $\ell \bar \ell$. The spin state $\ket{s_\ell s_{\bar \ell}}$ is then, in general, not normalised, but it contributes to $\ket{\psi}$ weighted by the $S$ matrix element $\mathcal{M}_{s_\ell s_{\bar \ell}}$. Consider the simple case where $\ket{\psi}$ is the following pure, unnormalised state:
	\begin{equation}
		\ket{\psi} =  a\left(  \ket{++} + \ket{--} \right)+ b \left( \ket{+-} - \ket{-+} \right)\, ,
		\label{eq:psi1}
	\end{equation}
	where we have quantised the spins along the $z$ axis. We recognise this state to be a linear combination of a triplet and singlet state. Given that these two states are orthogonal, they do not interfere in the modulus squared of the final state:
	\begin{equation}
		\langle \psi | \psi \rangle = \sum_{s_{\ell}s_{\bar \ell}}|\mathcal{M}_{s_{\ell}s_{\bar \ell}}|^2 = |a|^2+|b|^2\, .
		\label{eq:ModSq}
	\end{equation}
	We can consider also spin observables by insertion of the spin operators $S^\ell_i = \sigma_i \otimes \mathds{1}_2,\, S^{\bar\ell}_i =  \mathds{1}_2 \otimes\sigma_i$, with $\sigma$ the Pauli matrices. The correlation between the two spins, both measured along the $z$ axis, is given by 
	\begin{equation}
		\left\langle \psi \middle| S_z^\ell S_z^{\bar \ell} \middle| \psi \right\rangle= |a|^2-|b|^2\, ,
		\label{eq:ZCorr}
	\end{equation}
	which, like the modulus squared, only depends on the two coefficients squared, so it receives no contributions from interference. It would be too hasty to conclude at this point that the orthogonality of the two states prevents them from interfering in all observables. Indeed, interference generates the polarisation of $\ell$ along the $y$ axis, because the $S_y^\ell$ operator causes a spin flip:
	\begin{align}
		\left\langle \psi \middle| S_y^\ell \middle | \psi \right\rangle &= \langle \psi | \Big{(}a\left( |-+\rangle - |+-\rangle \right) + b\left( |--\rangle + |++\rangle \right) \Big{)} \\
										 &= 4\Im(ab^*)\, . \nonumber
										 \label{eq:YPol}
	\end{align}
	And similarly we find a nonzero interference when considering the expectation value of $S_z^\ell S_x^{\bar \ell}$, which gives the correlation between the two spins when measured along the $z$ and $x$ axis, respectively: 
	\begin{equation}
		\left\langle \psi \middle| S_z^\ell S_x^{\bar \ell} \middle | \psi \right\rangle = -2\Re(a b^*)\, .
		\label{eq:CorrZC}
	\end{equation}
	Measuring the full set of spin observables, as well as the modulus square, allows us to fully reconstruct $|\psi \rangle $. In contrast, a measurement of the modulus square alone cannot access the relative phase between $a$ and $b$. This is the essence of a process called ``quantum tomography''~\cite{Dariano:2003ngl}; we discuss quantum tomography at colliders in \cref{sec:Tomography}.

	Let us now add the effect of classical uncertainty, to describe polarisation and other degrees of freedom. This is best done using a density matrix. We recall that for a given state $\ket{\psi}$, which is represented by a statistical ensemble of pure states $\ket{\phi_i}$, the density matrix $\rho$ is the weighted sum of the projectors over $\ket{\phi_i}$:
	\begin{equation}
		\rho = \ket{\psi}\bra{\psi} = \sum_i p_i \ket{\phi_i}\bra{\phi_i}\, ,
	\end{equation}
	where $\{\ket{\phi_i}\}_i$ is a basis for the Hilbert space of the system, and the $p_i$ sum to one. In the present case of a bipartite qubit, $\rho$ will be a $2^2=4$-dimensional square matrix. 

	We will assume the initial state is an unpolarised beam, represented as an equiprobable ensemble $\rho_i$:
	\begin{equation}
		\rho_i = 1/4 
		\begin{pmatrix}
			1&0&0&0\\
			0&1&0&0\\
			0&0&1&0\\
			0&0&0&1\\
		\end{pmatrix}
		\, . 
		\label{eq:rhoi}
	\end{equation}
	We will have the scattering matrix $S$ act on the pure states as:
	\begin{equation}
		\begin{cases}
			| ++ \rangle \to \sqrt2 a|++\rangle\\
			| -- \rangle \to \sqrt{2}a|--\rangle\\
			| +- \rangle \to b\left(|+-\rangle + |-+\rangle\right) + c\left(|++\rangle + |--\rangle\right)\\
			| -+ \rangle \to b\left(|+-\rangle + |-+\rangle\right) + c\left(|++\rangle + |--\rangle\right)
		\end{cases}\, ,
		\label{eq:SMat}
	\end{equation}
	that is, in matrix form 
	\begin{equation}
		S = 
		\begin{pmatrix}
			\sqrt2a&0&0&0\\
			c&b&b&c\\
			c&b&b&c\\
			0&0&0&\sqrt2a\\
		\end{pmatrix}
		\, .
		\label{eq:SMat1}
	\end{equation}
	The action of $S$ was chosen to mimic the 4-fermion interactions we study in this paper. In particular, the action on the equal-sign states ($a$ channel) mimics the $\mathcal{O}_{\ell equ}^{1}, \mathcal{O}_{\ell equ}^{3}$ operators, while the action on the mixed-sign states ($b$ and $c$ channels) mimics the SM and the dipole operators, respectively. 

	The trace of the final state, which computes the weighted average of all the pure states square modulus is
	\begin{equation}
		A = \textrm{Tr}(S\rho_iS^\dagger)= |a|^2 + |b|^2 + |c|^2\, ,
		\label{eq:ModS}
	\end{equation}
	which does not receive contributions from interference. On the other hand, the correlation matrix of the two spins along the three axes is given by
	\begin{equation}
		\tilde{C} = \textrm{Tr}(\vec{S}^{\ell}\otimes\vec{S}^{\bar \ell} S \rho_i S^\dagger) =
		\begin{pmatrix}
			|b|^2+ |c|^2 & 0 & 0\\
			0 & |b|^2 - |c|^2 & 2\Im (b c^*)\\
			0 & 2\Im (b c^*) & |a|^2 - |b|^2 + |c|^2 
		\end{pmatrix}
		\, ,
		\label{eq:CorrS}
	\end{equation}
	which shows interference between the $b$ and $c$ channels in the $(yz)$ and $(zy)$ components. The $a$ channel, on the other hand, does not interfere with the other two channels because the initial states are in a classical admixture. If the $a$ channel had a nonzero action on the mixed-sign spin configurations, it would interfere with the $b$ channel, even though it induces an orthogonal spin configuration in the final state. 

	The simple examples in this subsection demonstrate how interference can be resurrected in spin correlations: The interference between orthogonal final states is generated by a spin-flip operator, but only if the states are in a quantum superposition.

	We now introduce the necessary formalism for describing quantum tomography measurements at a collider experiment.

	\section{Spin Correlations and Quantum Tomography}
	\label{sec:formalism}
	We study spin correlations in 4-fermion scattering $f \bar f \to \ell \bar \ell$. As argued before, the final state $\ket{\ell \bar \ell}$ can be expressed in terms of a $4 \times 4$ density matrix, called the spin-density matrix, defined using the $S$-matrix elements as follows~\cite{Afik:2022kwm}:
	\begin{align}
		R_{s_{\ell,1}s_{\ell,2},s_{\bar \ell,1}s_{\bar \ell,2}} & \equiv \frac{1}{N^2} \sum_{f,\bar f\rm \,d.o.f.} \mathcal{M}^*_{s_{\ell,2}s_{\bar \ell,2}}\mathcal{M}_{s_{\ell,1}s_{\bar \ell,1}} \, , \\
		\mathcal{M}_{s_{\ell}s_{\bar \ell}} & \equiv \bra{\ell(k_1,s_{\ell})\bar{\ell}(k_2,s_{\bar \ell}) }S{\ket{ f(p_1)\bar f(p_2)}}\, ,
		\label{eq:R}
	\end{align}
	where $\mathcal{M}_{s_{\ell}s_{\bar \ell}}$ are the polarised scattering matrix elements with $s_{\ell}$ ($s_{\bar \ell}$) the value of the $\ell$ ($\bar\ell$) spin along the quantization axis, taking values $+,-$. 

	\noindent		We average over discrete degrees of freedom (d.o.f.), such as spin  and color, by summing incoherently over them and dividing by an appropriate normalisation factor $N$ for each initial state fermion. The sum is incoherent because we assume the initial state is a mixed state with uniform statistical distribution over the initial d.o.f.~(i.e.~we assume unpolarised beams). This results in the final state being mixed~\cite{Afik:2022kwm}. In contrast, the final state spin d.o.f.~are in a quantum superposition.  Note also that while we defined $R$ to have four indices, the indices $(s_{\ell,1}s_{\bar \ell,1})$ and $(s_{\ell,2}s_{\bar \ell,2})$ can be combined to make $R$ a $4 \times 4$  matrix. 

	Any $4 \times 4$ matrix can be decomposed in terms of products of Pauli matrices $\sigma^i$ in terms of what are known as "Fano coefficients" as follows~\cite{Fano:1983zz}:
	\begin{align}
		R = 
		A \mathds{1}_4
		+\tilde{B}_i^+ \sigma^i\otimes \mathds{1}_2 + \tilde{B}_i^- \mathds{1}_2\otimes\sigma^j + \tilde{C}_{ij}\sigma^i \otimes \sigma^j \,,
	\end{align}
	where we can recognise the spin operator acting on a single lepton, $S_i^+ =  \sigma_i\otimes\mathds{1}_2, S_i^- = \mathds{1}_2\otimes \sigma_i$, as well as their product, $S_i^+S_j^- = \sigma_i\otimes \sigma_j$. The Fano coefficient can then be readily interpreted: $A$ is the overall normalisation of $R$, the vector $\tilde{B}_i^-$ $(\tilde{B}_i^+)$ quantifies the spin polarisation of the (anti-)lepton, while the matrix $\tilde{C}_{ij}$ contains the correlation between the spins of the two particles. The probability of a given spin configuration $(s_{\ell},s_{\bar \ell})$ being measured is
	\begin{equation}
		P(s_{\ell},s_{\bar \ell})= \frac{1}{N^2}\sum_{\rm f,\bar f\, d.o.f.} |\mathcal{M}_{s_{\ell}s_{\bar \ell}}|^2\, .
		\label{eq:PolProb}
	\end{equation}

	$A$ is the trace of $R$, that is the sum of all the probabilities. This is, of course, proportional to the differential cross section:
	\begin{equation}
		\frac{d\sigma}{d\Omega} = \frac{\alpha^2 \beta_{\ell}}{m_{\ell \ell}^2} A(m_{\ell \ell},\mathbf{k})\,,
		\label{eq:SigmaA}
	\end{equation}
	where $m_{\ell\ell}^2 = (p_{\ell}+p_{\bar \ell})^2$ is the di-lepton invariant mass squared, $\alpha$ an appropriate coupling constant, $\beta_\ell = \sqrt{1-4m_{\ell}^2/(m_{\ell \bar \ell}^2)}$ is the $\ell$ velocity and $\Omega$ the solid angle defined by the direction $\mathbf{k}$. 
	On the other hand, the probability of the spin measurement on $\ell$ having the outcome $(s_{\ell})$ can be obtained from the reduced density matrix, defined by tracing out the $\bar \ell$ spin d.o.f.:
	\begin{equation}
		\textrm{Tr}_{\bar \ell}(R)_{s_{\ell}s_{\ell}'} = \sum_{s_{\bar \ell}}\mathcal{M}_{s_{\ell}' s_{\bar \ell}}^*\mathcal{M}_{s_{\ell} s_{\bar \ell}}\, ,
		\label{eq:ProbPol1}
	\end{equation}
	\begin{equation}
		P(s_{\ell}) = \frac{1}{N^2}\sum_{ \rm f,\bar f,\, d.o.f.}|\sum_{s_{\bar \ell}} \mathcal{M}_{s_{\ell}s_{\bar \ell}}|^2 = \sum_{s_{\bar \ell}} P(s_{\ell},s_{\bar \ell}) + 2\Re(\mathcal{M}_{s_{\ell},-}\mathcal{M}_{s_{\ell},+}^*).
		\label{eq:ProbPol2}
	\end{equation}
	As we saw in the previous subsection, the probabilities of the spin measurements on a single particle, and thus also their correlations,  depend not only on the matrix element squared, but also on the interference between matrix elements with different spin structures. In contrast, the initial d.o.f.~are in a classical admixture and do not interfere. 

	At this point, one can normalise the spin-density matrix as  $\rho=R/\textrm{Tr}[R]$ to obtain a proper density matrix. We can now expand $\rho$ in the same form as the spin-production matrix (note the absence of tildes)
	\begin{align}
		\rho 
		=\frac{1}{4}(\mathds{1}_4
		+{B}_i^+ \sigma^i\otimes \mathds{1}_2 
		+ {B}_i^- \mathds{1}_2\otimes\sigma^i 
		+ {C}_{ij}\sigma^i \otimes \sigma^j) \, .
	\end{align}

	If the initial state is composite, like at a $pp$ collider, the full density matrix can be factorised as a classical ensemble of the individual partonic contributions, weighted by the luminosity function $I_q$~\cite{Afik:2020onf}:
	\begin{equation}
		\rho^{pp}_{ijkl} = \dfrac{\sum_q I_{q} R_{ijkl}^{qq}}{\sum_q I_{q} A^{qq}}\, .
		\label{eq:rhopdf}
	\end{equation}
	The luminosity functions are defined as:
	\begin{equation}
		I_{qq}(\hat m) = \frac{2 \hat m}{\sqrt{s}}\int_{\hat m}^{1/\hat m}\frac{dz}{z}q_q(\hat m z) q_{\bar q}\left(\frac{\hat m}{z}\right),
		\label{eq:LumFct}
	\end{equation}
	with $\hat m = m_{\ell \bar \ell}/\sqrt{s}$, and $q_q(x)$ is the PDF of the parton $q$. The numerical value of the PDFs are taken from the PDF4LHC collaboration~\cite{PDF4LHCWorkingGroup:2022cjn} and evaluated with the help of the Mathematica interface ManeParse~\cite{ManeParse}.  We will consider only $q = u,d,s$ because the luminosity functions of heavier quarks are negligibly small.	

	Finally, we will be interested in the density matrix integrated over the solid angle $d\Omega =  \sin \theta d\theta d \phi \equiv dz d\phi$, with $\Omega$ the scattering angle of the hard scattering, so we define an integrated spin-production matrix and density matrix:
	\begin{align}
		R^{\rm Integrated} =\frac{1}{4 \pi} \int dz d\phi R \, , \\
		\rho^{\rm Integrated} = \frac{R^{\rm Integrated}}{A^{\rm Integrated}}\, .
		\label{eq:rhoInt}
	\end{align}
	Having defined the necessary density matrix, we consider how to define the amount of entanglement it contains.

	\subsection{Entanglement Measures and Inequalities}
	\label{subsec:Entanglement_Inequalities}
	A density matrix represents an entangled state if and only if its density matrix cannot be expressed as a convex combination of product states, i.e.
	\begin{equation}
		\rho = \sum_{ij}p_{ij}\rho_i^+\otimes\rho_j^-\, ,
		\label{eq:entangrho}
	\end{equation}
	where $\rho_i^\pm$ is a one-qubit state density matrix on the Hilbert space of the (anti-)particle and the real-valued $p_{ij}$ are non-negative and sum to one.
	If such a decomposition does not exist, the state is said to be entangled~\cite{Schroedinger_1935}, because knowledge about the Hilbert space of each single subsystems does not translate to knowledge about the full Hilbert space, a purely quantum phenomenon. This formal definition does not provide however a quantitative handle on the entanglement; moreover explicitly checking whether a state is separable involves proving a negative, a generically hard task.  For these reasons various quantities have been devised to measure entanglement (see e.g.~\cite{Horodecki_2009} for a review). The specific case of two-qubit bipartite states is the best-understood one, and a plethora of observables, known as ``entanglement witnesses'' with explicit expressions are available. To be an entanglement witness, a measure of entanglement must attain its minimum value (usually zero) for pure states and its maximum value at maximally entangled states (such as Bell pairs).

	\paragraph{Concurrence}
	One entanglement witness that is often used is the concurrence $\mathcal{C}[\rho]$~\cite{Wootters_1998}, defined in terms of the spin-flipped density matrix $\tilde{\rho}= (\sigma_2\otimes \sigma_2)\rho^*(\sigma_2\otimes\sigma_2)$. The decreasingly ordered eigenvalues $\lambda_i$ of the  matrix $\omega= \sqrt{\sqrt{\tilde{\rho}}\rho\sqrt{\tilde{\rho}}}$ or, equivalently, the square root of the eigenvalues of the matrix $\Omega= \rho\tilde{\rho}$, are used to define~\cite{Wootters_1998}
	\begin{equation}
		\mathcal{C}[\rho] = \textrm{max}(0,\lambda_1 - \lambda_2 - \lambda_3 - \lambda_4).
		\label{eq:Conc}
	\end{equation}
	Null concurrence is attained for separable states, while the maximum value of $1$ is attained for maximally entangled states.

	\paragraph{CHSH Inequality}
	Arguably the most famous form of non-classical correlations is the violation of a set of inequalities, the ``Bell inequalities'', which are always satisfied in local and realistic theories. A particularly useful form, the CHSH (Clauser-Horne-Shimony-Holt) inequality~\cite{CHSH_1969}, allowed the first-ever experimental determination of Bell non-locality~\cite{PhysRevLett.28.938} and is still a cornerstone of quantum information studies. 

	For two-qubit systems, a criterion due to Horodecki for violating the CHSH inequality can be expressed in terms of the spin correlation matrix $C$ only~\cite{Horodecki1995340}. One defines $(m_1,m_2,m_3)$ as the eigenvalues of $M=C^TC$ in decreasing order. According to the Horodecki criterion, the CHSH inequality is violated if and only if~\cite{Horodecki1995340}
	\begin{equation}
		\mathfrak{m}_{12} = m_1 + m_2 > 1\, .
		\label{eq:m12}
	\end{equation}
	Maximum violation is obtained at $\mathfrak{m}_{12}=2$, and separable states have $\mathfrak{m}_{12}=0$. It is worth noting that, in general, violation of Bell inequalities is a stronger condition than entanglement; the two notions coincide only for pure two-qubit bipartite states~\cite{Gisin1991201}. 
	\paragraph{Simplified Criteria}
	It is clear that, if the polarisation vectors $B^\pm$ are vanishing, all the information about entanglement is contained in the correlation matrix $C$. 
	This happens if the individual leptons are not produced in a polarised state by the interaction due to, for example, separate conservation of $C$ and $P$ parities~\cite{Bernreuther:2015yna}. 
	In that case, the $C$ matrix elements is also constrained to be block-diagonal~\cite{Afik:2020onf}. Sufficient (but not necessary) conditions for Bell inequality violation and entanglement can be obtained from the Peres-Horodecki criterion~\cite{Peres:1996dw} with reference only to the diagonal elements of the $C$ matrix. 
	For instance, for top pair production, the opening angle coefficient
	\begin{equation}
		D = \frac{1}{3}(C_{kk}+C_{rr}+C_{nn})\, , 
		\label{eq:simplified}
	\end{equation}
	is used in Ref.~\cite{Maltoni:2024tul}, together with other simple linear combinations that have different relative signs of the $C_{ii}$ coefficients.

	We will not pursue this direction, because as we showed in \cref{sec:ToyModel}, resurrection of interference is active only in the off-diagonal elements of the $C$ matrix. Moreover, EW interactions violate both $P$ and $C$ parities.

	\paragraph{Quantum Discord}
	It has long been recognised~\cite{Bennett:1998ev} that there exist separable mixed states whose correlation cannot be described by a classical theory. This has prompted the search for finer discriminant between classical and quantum states, see Ref.~\cite{bera2017quantum} for a recent review. Various measures have been devised, focusing on different aspects of quantum mechanics, such as the fact that a local measurement on subsystems can induce disturbances in the overall system. Most of the measures of quantum correlations are called ``discord'', because they measure the difference between two generalisations of spin correlations that would coincide in the classical case. Many of the measures of discord involve complicated minimisation processes and are hence not easily amenable to calculations, without employing simplifying assumptions on the nature of the state, as can be done for example in top pair production~\cite{Afik:2022dgh}. 

	In this work we will focus on one definition of quantum correlations that has a relatively simple expression, the Local Quantum Uncertainty (LQU)~\cite{Girolami:2012tz}. It quantifies the irreducible quantum uncertainty in local measurements on a subsystem. The LQU is zero if and only if there exists at least one observable that can be measured on a subsystem that is not affected by quantum uncertainty. The LQU is defined as~\cite{Girolami:2012tz}
	\begin{align}
		LQU(\rho) &= 1- \textrm{max}(\textrm{Eig}\left(W \right))\, ,\\
		W_{ij} &= \textrm{Tr}(\sqrt\rho(\sigma_i\otimes\mathds{1}_{2})\sqrt\rho(\sigma_j\otimes\mathds{1}_{2}))\, .
		\label{eq:LQU}
	\end{align}
	We now consider how to experimentally measure spin polarisations and correlations.
	\subsection{Tomography}\label{sec:Tomography}

	Collider experiments are generically not able to directly measure the spin of outgoing particles. Luckily, due to angular momentum conservation, the spin information of a decaying particle is imprinted in its decay products. By selecting one of the decay products $d^- (d^+)$ for the outgoing $\ell$($\bar\ell$), we can relate the spin density matrix to the doubly differential distribution of the cross section in the decay product angles~\cite{Baumgart:2012ay}. By partially integrating over the differential cross section, one obtains~\cite{Bernreuther:2004jv,Bernreuther:2015yna}
	\begin{align}
		\frac{1}{\sigma}	\frac{d\sigma}{d\cos\theta_{\pm}^i} & = \frac{1}{2} \left(1\pm \alpha_{\pm} B_{i}^{\pm}\cos\theta_{\pm}^i\right)\, ,\\
		\frac{1}{\sigma}		\frac{d\sigma}{d(\cos\theta_{+}^i\cos\theta_-^j)} &= -\frac{1}{2} \left(1 + \alpha_{+}\alpha_{-}C_{ij}\cos\theta_{+}^i\cos\theta_{-}^j\ln(|\cos\theta_{+}^i\cos\theta_{-}^j|)\right)\; ,
		\label{eq:Tomography}
	\end{align}

	where we define the polar angle with respect to a given direction $\hat{u}_i$ through $\cos(\theta_\pm)^i\equiv \hat{d}^\pm\cdot\hat{u}_i$, with $\hat{d}^-(\hat{d}^+)$ the direction of flight of the decay product we picked for the lepton (antilepton). The coefficient $\alpha_\pm$ is called the ``spin-analysing power'', and it depends on the specific decay chain that led to the decay product $d^\pm$; numerical values of the spin-analysing power of various $\tau$ decay channels can be found in Ref.~\cite{Bernreuther:2021elu}.

	From \cref{eq:Tomography} it follows that the spin correlations can be obtained by averaging the angular distributions~\cite{Bernreuther:2004jv,Bernreuther:2015yna}:
	\begin{align}
		B_{i}^\pm &= 3\frac{<\cos\theta_\pm^i>}{\alpha_\pm}\\
		C_{ij} &= 9\frac{<\cos\theta_+^i\cos\theta_-^j>}{\alpha_+\alpha_-}.
		\label{eq:averages}
	\end{align}

\Cref{eq:Tomography} is the basis of \textit{quantum tomography}: from repeated measurement of angular correlations, one can reconstruct the density matrix, hence the state, of the outgoing $\ell \bar \ell$ pair. We note that \cref{eq:averages} is not the only way to extract the Fano coefficients from angular correlations, but all approaches are based on \cref{eq:Tomography} and they perform similarly in terms of experimental uncertainties~\cite{Altakach:2022ywa,Ehataht:2023zzt}. 

	We compute the density matrix in the helicity basis $\{\hat r,\hat n,\hat k\}$, where $\hat k$ points along the $\bar \ell$ 3-momentum and the other two unit vector are defined with reference to a fourth unit vector $\hat p$, which we take to be the beam axis: 
	\begin{align}
		\hat p \cdot \hat k &= z\, ,\\
		\hat r &= \frac{1}{\sqrt{1-z^{2}}}(\hat p - z\hat k)\, ,\\
		\hat n &= \frac{1}{\sqrt{1-z^{2}}}\hat k \times \hat p\, .
		\label{eq:HelBasis}
	\end{align}
	The helicity basis is event-dependent, thus the state that one reconstructs is in general not a proper quantum state but what is known as a ``fictitious state''~\cite{Afik:2022kwm}.       
Nonetheless, fictitious states retain the information of the single measurements that were used for its reconstruction, even though this information might get smeared~\cite{Cheng:2023qmz}.

	We should note at this point that measuring the angular correlations of the decay products of $\ell$ requires a reconstruction of the momentum of $\ell$, as well of the c.o.m.~frame. At lepton colliders, the center of mass is known with precision, and possible invisible particles in the decay products can be reconstructed using impact parameters~\cite{Jeans:2015vaa,Altakach:2022ywa}. The situation is more complicated at hadron colliders, where the composite nature of the hadron makes the c.o.m.~frame of the hard scattering \textit{a priori} unknown. We will revisit this issue later.

	Based on the discussion in this section, we want to highlight a conceptual difference between the existing descriptions of interference resurrection in the literature and the mechanism presented in \cref{sec:ToyModel}. In diboson production, interference resurrection has been studied extensively, both in the triple-W~\cite{Helset:2017mlf,Azatov:2019xxn,ElFaham:2024uop} and triple-gluon~\cite{Dixon:1993xd} contexts. It has been noted that the operator $\mathcal{O}_{3V}$ does not interfere with the Standard Model (SM) in the process $f \bar f \to V V$~\cite{Simmons:1989zs,Simmons:1990dh}, but interference is restored by considering a larger process where $f\bar f \to V V$ is an intermediate stage, $VV$ being off-shell~\cite{Dixon:1993xd,Azatov:2017kzw,Aoude:2019cmc}. In density matrix terminology, Ref.~\cite{Aoude:2023hxv} discusses resurrection in electroweak boson production. It has also been observed that interference can be artificially canceled if a strict narrow-width approximation is used~\cite{Helset:2017mlf}, making a correct quantum treatment of the intermediate $VV$ state crucial. As reviewed, measuring spin polarization typically requires the decay of the final-state particle. However, our discussion in \cref{sec:ToyModel} clarifies that the key factor is not decay or the particle being off-shell, but a proper quantum mechanical treatment. Thus, if we had direct access to the spins of (on-shell) particles at colliders, quantum mechanics would still govern the resurrection phenomenon, independent of the specific method to measure spin.

	Having defined the observables needed for the measurement of spin observables, we turn now to their calculation, and to the question of whether they can display resurrection of interference. First, we categorise the SM and SMEFT contributions to $f\bar f \to \ell \bar \ell $.
	\subsection{Resurrection of the Dipole}
	We are now ready to discuss the resurrection of dimension-6 operators in 4-fermion scattering. As we noted before, only the left-right operators $\mathcal{O}^{1}_{\ell e q u}, \mathcal{O}^{3}_{\ell e q u}$ and the dipole operators $\mathcal{O}_{V\ell}$ exhibit different helicities structure than the SM. It is clear from the discussion in \cref{sec:ToyModel} that the contact operators do not interfere with the SM at all, even in the off-diagonal elements of the $R$ matrix, because we sum over (fixed) initial fermion polarisation $a,b$. So we will not consider them going forward.

	We turn now to the dipole operator class. As we just argued, modifying the helicity selection rules for the initial states makes the interference impossible; therefore we focus on final-state dipole operator $\mathcal{O}_{V\ell}$. Just like in the SM, the gauge boson is fixed to be transverse by the vanishing mass of initial fermions, regardless of any helicity selection rules involving the final leptons. The dipole operator imposes the $(T++), (T--)$  configuration in the chiral limit,
	which means that the full process $f\bar f \to V \to \ell \bar \ell $ will have the $(+-++),(+---),(-+++),(-+--)$ structure in the chiral limit:
	\begin{equation}
		\mathcal{M}_{s_{\ell}s_{\bar \ell}}^{\textrm{Dipole}} = 0 + \mathcal{O}(\frac{m_l}{m_{ll}})\quad \textrm{if} \; s_{\ell} \neq s_{\bar \ell}\, .
		\label{eq:HelDip}
	\end{equation}
	Unlike the contact operators, then, the dipole operator interference with the SM populates the off-diagonal entries of the $R$ matrix. The leptonic dipole operators are unique since they are the only class of operators suffering from helicity suppression in the interference, but still contributing an interference term to spin correlations. 

	\subsection{Resurrection, Truncation and Unitarity}
	Before proceeding, it is important to address a subtlety related to the mismatch in large-energy behaviour of the various entries of the $R$ matrix, which arises due to the noninterference and resurrection phenomena: the $R$ matrix, truncated at order $\Lambda^2$, violates perturbative unitarity at energies lower than the cutoff.
	To clarify, let us separate the various contributions to a given observable based on their $\Lambda$ dependence and energy scaling from naive dimensional analysis (NDA). The dipole operator has the NDA scaling of $vm_{\ell\ell}/\Lambda^2$ for interference contributions. However, due to the helicity selection rule only the SMEFT squared scales with energy, while the interference is proportional to the constant chirality-flip factor $vm_\ell/\Lambda^2$~\cite{Haisch:2023upo}:
	\begin{equation}
		\frac{d \sigma}{d\Omega} \overset{m_{\ell\ell}\gg m_\ell}{\to} \frac{d \sigma}{d\Omega}^{(0)} +\frac{vm_{\ell}}{\Lambda^2} \frac{d \sigma}{d\Omega}^{(2)}+ \frac{v^2m_{\ell\ell}^2}{\Lambda^4} \frac{d \sigma}{d\Omega}^{(4)} \, ,
		\label{eq:SigmaHighE}
	\end{equation}
	and similarly we can define $A^{(n)}$ and $R^{(n)}$ by looking at the high-energy behaviour.
	This means that $A$ receives a small, energy-independent contribution at order $\Lambda^2$. However, due to resurrection, the off-diagonal entries of the $R$ matrix, at order $\Lambda^2$, have the NDA scaling of $\frac{m_{\ell\ell}v}{\Lambda^2}$ at large energies: 
	\begin{align}
		R_{ii} &\overset{m_{\ell\ell}\gg m_\ell}{\to} R_{ii} ^{(0)} + \frac{vm_\ell}{\Lambda^2}R_{ii}^{(2)} + \frac{v^2 m_{\ell\ell}^2}{\Lambda^4}R_{ii}^{(4)}\, ,\\
		R_{ij} &\overset{m_{\ell\ell}\gg m_\ell}{\to} R_{ij} ^{(0)} + \frac{v m_{\ell\ell}}{\Lambda^2}R_{ij}^{(2)}+\frac{v^2 m_{\ell\ell}^2}{\Lambda^4}R_{ij}^{(4)}\; ,\, i\neq j\, .
		\label{eq:RHighE}
	\end{align}
	Therefore, if we neglected the SMEFT squared contributions, the off-diagonal terms of the normalised density matrix would also grow like $m_{\ell\ell}$ at large energies:

	\begin{equation}
		\frac{R_{ij}^{(0)} + \frac{v m_{\ell\ell}}{\Lambda^2}R_{ij}^{(2)}}{A^{(0)} + \frac{v m_{\ell}}{\Lambda^2}A^{(2)}} \overset{m_{\ell\ell}\gg m_\ell}{\to} \frac{R_{ij}^{(0)} + \frac{v m_{\ell\ell}}{\Lambda^2}R_{ij}^{(2)}}{A^{(0)}}\quad i \neq j \, .
	\end{equation}
	If we instead include the order $\Lambda^4$ contributions, both the diagonal and off-diagonal entries of $R$ grow like $m_{\ell\ell}^2$ at large energies. Then, the energy dependence asymptotically cancels and the $\rho$ matrix elements approach constant values at large energies. 

	Truncating at order $\Lambda^{2}$ would result in a violation of perturbative unitarity in the $\rho$ matrix, at energies lower than the nominal cutoff scale $\Lambda$. Such a truncation is, however, clearly not permitted, because the SMEFT squared contribution dominates over the interference. Consequently, this affects the entanglement witnesses, causing them to grow over their theoretical maximum of 2 for $\mathfrak{m}_{12}$ and 1 for $\mathcal{C}[\rho]$. The sensitivity of quantum information observables to unitarity violations has been explored in~\cite{Aoude:2023hxv} for longitudinal boson scattering. We emphasise that such a violation of perturbative unitarity happens at energies lower than $\Lambda$, and as such does not seem directly related to possible violation of unitarity at or past the cutoff; moreover the full $R$ matrix up to order $\Lambda^4$ does not exhibit violation of perturbative unitarity at any energy, even past the cutoff.

	In order to avoid spurious violations of unitarity while still being able to observe the effect of resurrection of the interference, we define $\rho^{\textrm{No Int}}$ by ignoring the interference terms:
	\begin{equation}
		\rho^{\textrm{No Int}} = \frac{R^{(0)} + (\frac{v m_{\ell\ell}}{\Lambda^2})^2R^{(4)}}{A^{(0)} + (\frac{v m_{\ell\ell}}{\Lambda^2})^2A^{(4)}} \, .
		\label{eq:RhoNoInt}
	\end{equation}
	Another option would be to normalise by the full cross-section $A$. While at large energies the two choices differ negligibly, at low energies the density matrix would not be properly normalised; thus we find the definition in \cref{eq:RhoNoInt} more appropriate.

	In the absence of resurrection, the difference between $\rho^{\textrm {No Int}}$ and the full spin density matrix is proportional to the small, energy-independent ratio $vm_\ell/\Lambda^2$. On the other hand, resurrection of interference causes the difference between $\rho$ and $\rho^{\textrm{No Int}}$ to grow linearly with energy. A successful resurrection should also improve the sensitivity at moderate and low energies.

	We now explore the NP reach and resurrection properties of spin correlation and compare them to the reach of the quantum information observables.

\section{Case Study: $\tau$-Dipoles at Colliders} \label{sec:results}
	Having motivated the choice for the class of dipole operators, we will specialise to $\tau$ leptons. The motivation is twofold: first, flavour-universal BSM models, as well as loop effects in the SM, give rise to larger contributions to the $\tau$ dipole than to the other lighter fermions~\cite{eidelman2007theory}.~\footnote{Such enhancements can be even larger in scenarios with new bosons coupled to both lepton chiralities, yielding contributions to dipoles $\propto m_\ell m_F/\Lambda^2$, where $m_F$ denotes the mass of the heavy fermion running in the loops~\cite{Feruglio:2018fxo}. Examples are leptoquark models~\cite{Bernreuther:1996dr,Dorsner:2016wpm}, where $F$ is a heavy quark such as the bottom-quark~\cite{Cheung:2001ip,Dorsner:2019itg}, or scenarios where $F$ is a heavy vector-like lepton~\cite{Crivellin:2021rbq}.} Second, unlike the electron and muon $g-2$, which have been measured to incredible precision~\cite{Aoyama:2020ynm,Muong-2:2021ojo}, direct low-energy measurements of $g_\tau -2$ are precluded by the tiny lifetime of the $\tau$.  Current and prospective sensitivity to tau dipoles comes from high-energy collisions: a direct measurement at a fixed-target experiment has been proposed~\cite{Fomin:2018ybj,Fu:2019utm}, while indirect bounds are extracted from collider experiments such as $B$-factories~\cite{Bernabeu:1994wh,Bernabeu:2007rr,Bernabeu:2008ii,Eidelman:2016aih,Rajaraman:2018uyb,Belle:2021ybo}, electron-electron~\cite{Gonzalez-Sprinberg:2000lzf,DELPHI:2003nah}, proton-proton~\cite{Galon:2016ngp,Haisch:2023upo}, or ion-ion~\cite{ATLAS:2022ryk,CMS:2022arf,Verducci:2023cgx} collisions, by constraining the anomalous $\tau$-lepton coupling to off-shell photons or detecting a radiation zero in radiative $\tau$ decays~\cite{Laursen:1983sm,Rodriguez:2003tt}.
	All these probes suffer from non-interference and, partly as a result, the current sensitivity is orders of magnitude above the precise SM prediction~\cite{Fael:2013ij}. Moreover the sign and phase of the Wilson coefficients remain unconstrained. 

	Therefore, in this work we will consider the following lagrangian:
	\begin{align}
		\mathcal{L} &= \mathcal{L}_{\mathrm{\textrm SM}} + \mathcal{L}_{\mathrm{\textrm SMEFT}} \, ,\\
		\mathcal{L}_{\mathrm{\textrm SM}} &\supset eQ_{\ell}A_\mu \bar \ell\gamma^\mu \ell + \frac{e}{s_Wc_W}\left( Z_\mu \bar \ell \gamma^\mu(Q_{V\ell} +Q_{A\ell}\gamma^5)\ell \right) \, ,\\
		\mathcal{L}_{\mathrm{\textrm SMEFT}} &=
		\dfrac{c_{\tau B}}{\Lambda^2} \Big{(}\overline{L}_{L}\sigma^{\mu\nu} \tau_{R}\Big{)} H B_{\mu\nu} + \dfrac{c_{\tau W}}{\Lambda^2} \Big{(}\overline{L}_{L}\sigma^{\mu\nu} \sigma_3 \tau_{R}\Big{)} H W_{\mu\nu}^3 + \mathrm{h.c.} \; ,
		\label{eq:SMEFTDipole}
	\end{align}

	\noindent where $L_L= (\nu_{\tau L},\,\tau_L)^T$ is the left-handed third-generation lepton doublet, $\tau_R$ is the right-handed third generation weak singlet, $H$ is the Higgs boson, and $B^{\mu\nu}$ and $W_{\mu\nu}^{3}$ stand for the $U(1)_Y$ and the neutral $SU(2)_L$ field-strength tensors, respectively. After electroweak-symmetry breaking, we can rotate the effective couplings by the weak mixing angle, 
	\begin{align}
		c_{\gamma} &= \cos \theta_W \,c_{\tau B} - \sin \theta_W \,c_{\tau W}\, , \\
		c_{Z}     &=  \sin \theta_W \,c_{\tau B} + \cos \theta_W \,c_{\tau W}\, .
		\label{eq:SMEFTDipole-bis}
	\end{align}

	\noindent We further set the Higgs field to its vacuum expectation value  $v$, which renders the dimension-6 operators a three-point interaction. The rotated couplings in \cref{eq:SMEFTDipole-bis} then correspond to the $\tau$-lepton dipole interactions with the photon and the $Z$-boson. Neglecting higher-order corrections, it is possible to relate the effective coupling $c_{\gamma}$ to contributions to the anomalous magnetic and electric moments of the $\tau$-lepton,
	\begin{align}
		\Delta a_\tau &= \dfrac{2\sqrt{2}}{e} \dfrac{m_\tau v}{\Lambda^2} \Re(c_{\gamma})+\dots\, ,\\
		\Delta d_\tau &= -\sqrt{2} \dfrac{ v}{\Lambda^2} \Im(c_{\gamma})+\dots\,,
		\label{eq:AnMoms}
	\end{align}

	\noindent where the dots stand for loop corrections, cf.~Ref.~\cite{Aebischer:2021uvt,Feruglio:2018fxo}. Analogously, anomalous weak magnetic and electric moments $\Delta a_{\tau}^Z,\,\Delta d_{\tau}^Z$ can be defined. We will also find it useful to define $C_{\gamma(Z)}= c_{\gamma(Z)}/\Lambda^2$.

	In the rest of this section we present our results.  We analytically calculate the $R$-matrix at the level of undecayed $\tau$ leptons in the EW SM at tree level, as well as with the addition of the SMEFT dipole operators $\mathcal{O}_{\gamma\tau}, \mathcal{O}_{Z\tau}$ with the help of FeynArts and FormCalc~\cite{Hahn:1998yk,Hahn:2000kx}. The computation of the $R$-matrix is not novel, but we collect the coefficient of the resulting Fano decomposition in \cref{sec:Fano} for the reader's convenience.

	After integrating over $z$, we calculate, as a function of $m_{\tau\tau}$, the Bell inequality marker $\mathfrak{m}_{12}$, the concurrence $\mathcal{C}[\rho]$, and the local quantum uncertainty LQU\@. We pick as a benchmark point the current best constraint $|C_{\gamma}| \leq (1.5~\textrm{TeV})^{-2}$~\cite{Haisch:2023upo} and, since our goal is to compare the reach of different observables to NP, we apply it to $|C_{Z}|$ as well. We consider 4 possible phases for the Wilson coefficient, to showcase the sensitivity of different observables to the phase of the Wilson coefficients. We comment on the deviation of these observables when the SMEFT operators are turned on, which suggests that single elements of the $C$ matrix have  better sensitivity than quantum information observables.

Finally, we compute the sensitivities, and confirm that single elements of the $C$ matrix have better sensitivity both at high and low energies to the SMEFT operators, as well as better discrimination power between different phase choices for the Wilson coefficient, hence to the CP properties of the UV theory. For $e e$ colliders, the sensitivities are computed including the effects of tau decays, using MadGraph Monte Carlo (MC) simulations, both with and without detector and initial-state radiation (ISR) effects. We also comment on the possible sensitivities of $p p$ colliders.

\subsection{Quantum Observables and Spin Correlations}
\label{sec:colliders}
\subsubsection{Lepton Colliders}

\begin{figure}[ht]
	\centering
	\includegraphics[width=1.2\textwidth]{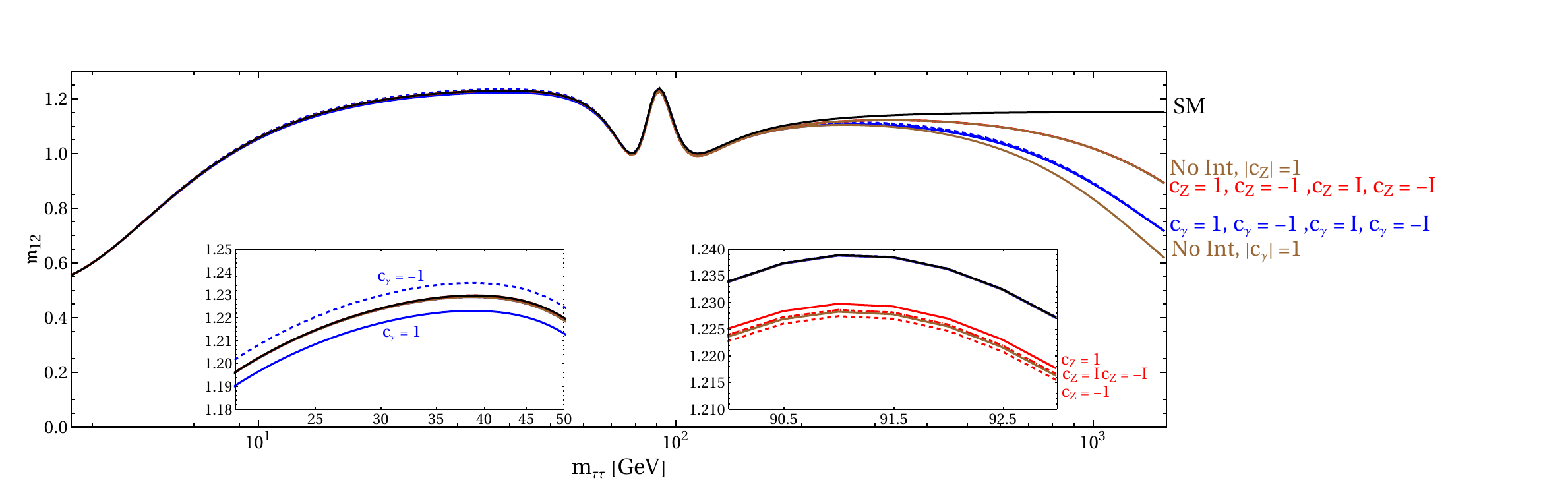}
	\caption{The $\mathfrak{m}_{12}$ Bell inequality marker for $ee\to \tau\tau$, $ C_{\gamma/Z}= 1/1.5\, \mathrm{TeV}^{-2}$. The insets show the local maxima of entanglement to appreciate the deviations due to the photon dipole operator (around $m_{\tau\tau} = 40$ GeV) and $Z$ dipole operator (around $m_{\tau\tau} = m_Z$). For $m_{\ell\ell}\gg m_Z$, the SM approaches a constant value, while the SMEFT operators bring down the entanglement of the $\tau$ pair state. Resurrection of interference of the photon dipole can be observed, but the interference and SMEFT squared contributions partially cancel each other. Moreover, while at the local maxima the $c= 1$ and $c= -1$ contributions can be distinguished, this is not true for $m_{\ell\ell}\gg m_Z$; imaginary Wilson coefficients, on the other hand, contribute negligibly at low energies and cannot be distinguished from real coefficients at large energies. }
	\label{fig:m12e}
\end{figure}

\begin{figure}[t]
	\centering
	\includegraphics[width=1.2\textwidth]{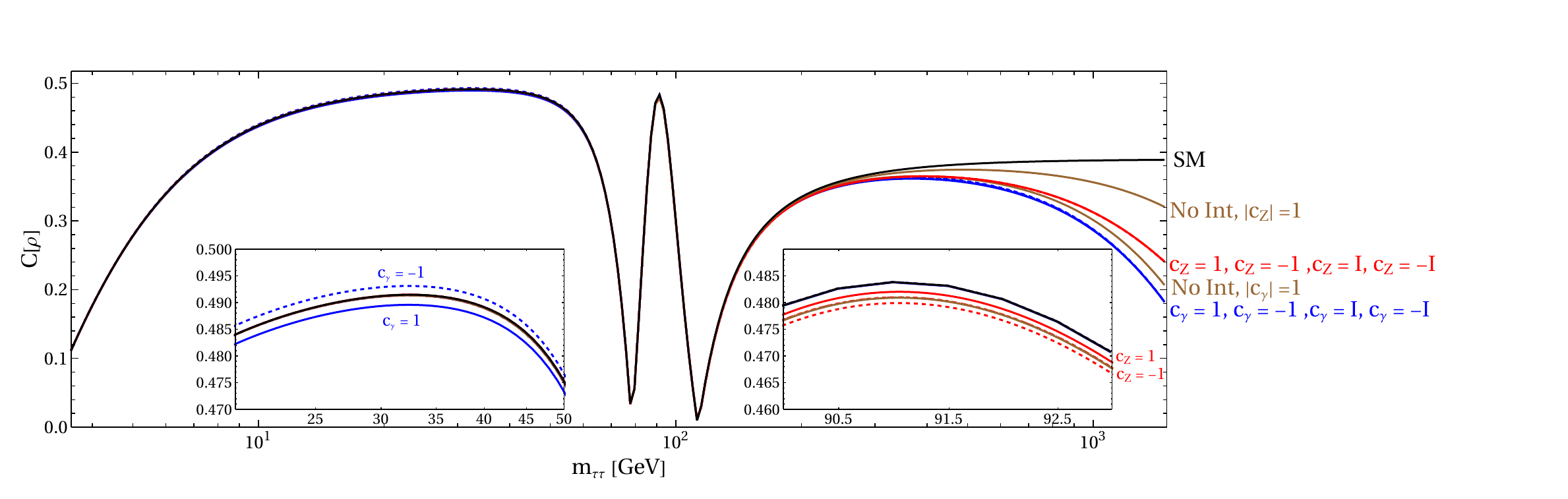}
	\caption{Same as \cref{fig:m12e}, but here the Concurrence $C[\rho]$, which quantifies entanglement directly, is plotted. In this case, resurrection of interference can be observed for both the photon and $Z$ dipole operators, and the interference and SMEFT squared contributions both decrease entanglement. However, at large energies the sign and phase of the Wilson coefficient cannot be determined, and at low energies there is limited sensitivity to the imaginary part of the Wilson coefficients.}
	\label{fig:conce}
\end{figure}
\begin{figure}[t]
	\centering
	\includegraphics[width=1.2\textwidth]{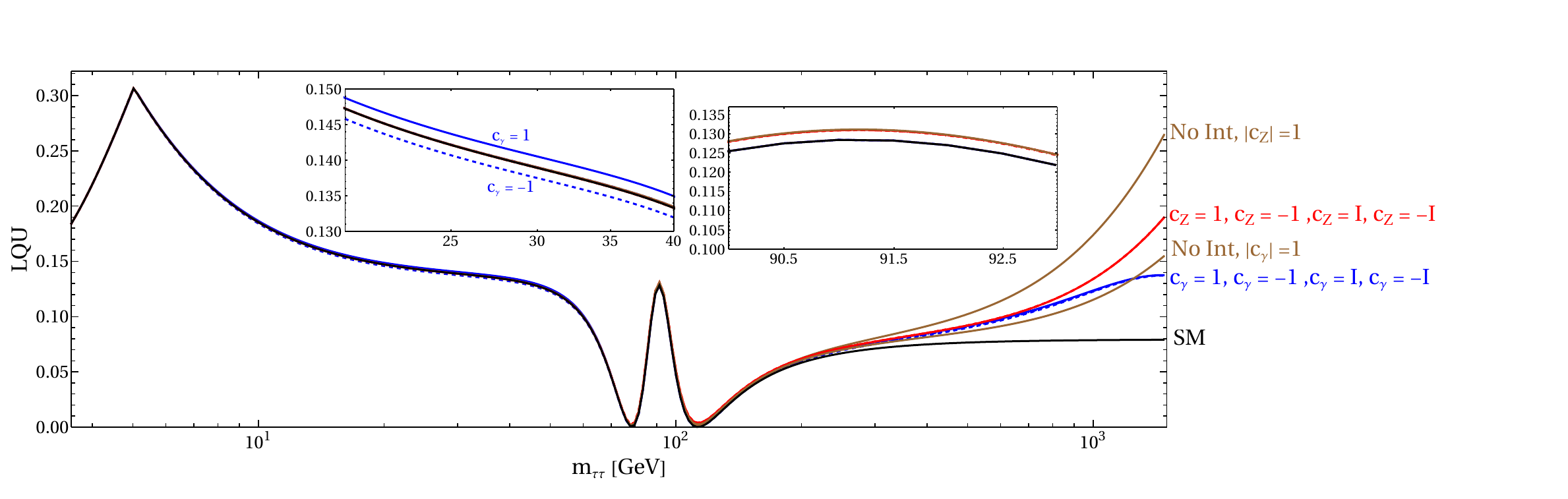}
	\caption{Same as \cref{fig:m12e}, but for the LQU, which quantifies quantum discord. The conclusions are mostly similar as in \cref{fig:m12e}, but resurrection of interference can be observed for the Z dipole, as well.}
	\label{fig:lque}
\end{figure}
\begin{figure}[t]
	\centering
	\includegraphics[width=1.2\textwidth]{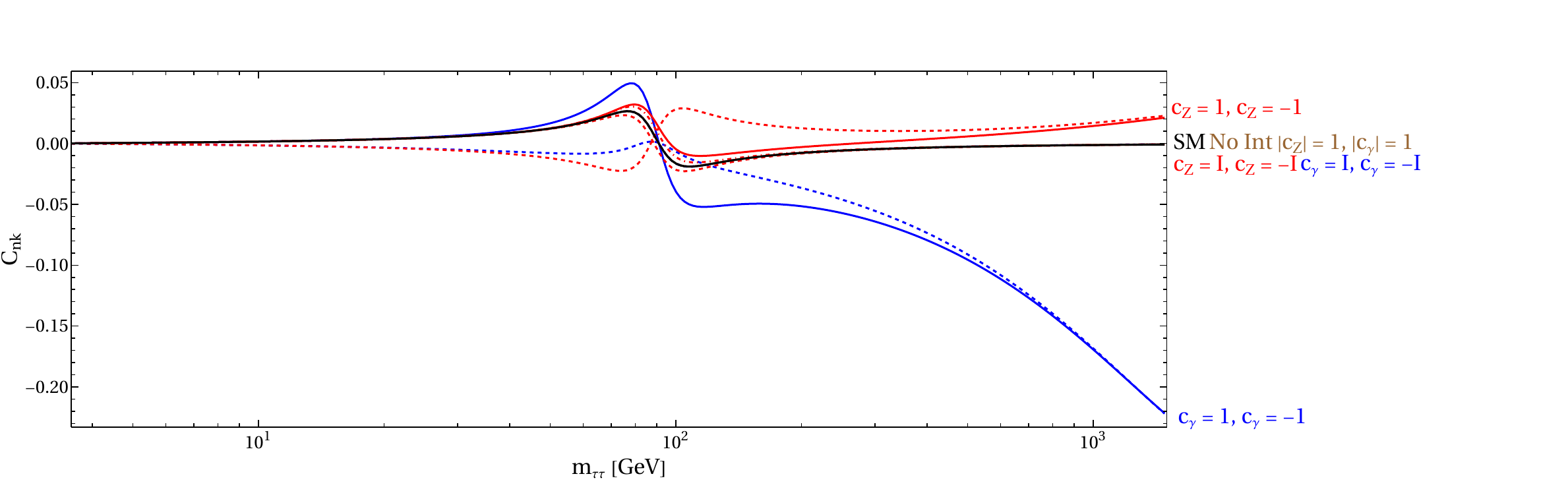}
	\caption{Same as \cref{fig:m12e}, but for the $(nk)$ element of the spin correlation matrix. Unlike the quantum information observables, here the SMEFT squared contribution is negligible, and the interference gives the leading deviation from the SM. Moreover, the sign of the real part of the Wilson coefficient changes dramatically the shape of the curve, everywhere but near the NP scale $\Lambda$. Note that the imaginary part of the Wilson coefficient contributes modestly to negligibly, but the roles of imaginary and real parts are exchanged in the $(nr)$ element, such that the phase of the Wilson coefficient can be obtained by examining all the elements of the $C$ matrix.}
	\label{fig:c23e}
\end{figure}
In \cref{fig:m12e,fig:conce,fig:lque,fig:c23e} we show the spin correlation observables for $ee\to \tau\tau$, at energies relevant for both past (LEP), current (Belle II), and proposed lepton colliders (FCC-ee, CLIC and ILC).  Of course, while we refer to a $ee$ collider, the results for a $\mu\mu$ collider would be exactly the same at the energies we consider here, because in both cases the mass of the initial leptons can be neglected.

In the SM, the maximum amount of entanglement can be observed at energies close to 30-50 GeV, where the photon exchange dominates and the external fermions can be considered as massless, and around the $Z$ pole, where the $Z$-exchange dominates. As a result, around these energies one can observe the largest deviations in the spin correlation observables: adding the $\gamma$ dipole operator changes the correlation pattern around the first maximum, while it has negligible effect at the $Z$ pole, and vice-versa adding the $Z$ dipole has a clearly observable effect at the $Z$ pole, being a resonant contribution there.

Observing \cref{fig:m12e}, one can notice how resurrection is successful in the Bell inequality marker $\mathfrak{m}_{12}$, in the sense that the interference has a non-negligible effect at large energies, but unfortunately the effect is to reduce the deviation from the SM, making the observable less sensitive compared to if resurrection did not occur. Moreover, at large energies the phase of the Wilson coefficient does not make any significant difference in the Bell inequality violation. Around the first maximum, however, the phase does play a role, with the pure imaginary coupling (hence $CP$-violating) benchmark increasing the amount of Bell inequality violation, and the opposite happening for pure real couplings ($CP$-conserving). 

We show the concurrence in~\cref{fig:conce}. We have a similar situation regarding the phase of the Wilson coefficient, but unlike the case of $\mathfrak{m}_{12}$, here the interference contribution are actually of the same sign as the SMEFT squared, allowing for increased sensitivity thanks to resurrection. 

Since the concurrence is a finer discriminant of quantum correlation than Bell inequality violation, one might imagine that the quantum discord, being an even finer discriminant, might fare even better as a NP probe. This is not the case, as can be seen in~\cref{fig:lque}: just like the case of $\mathfrak{m}_{12}$, the interference contributions for the LQU have opposite sign as the SMEFT squared contributions. It is interesting to note how the LQU features a maximum at small energies that is absent from the entanglement markers, as well as the fact that there is an increase in LQU when the SMEFT operators are added, while entanglement and Bell inequality violation fall off. 

Finally, we consider the elements of the $C$ matrix. In \cref{fig:c23e}, we plot the $(nk)$ element of the $C$ matrix. In the SM, this entry receives only a contribution proportional to the imaginary part of the $Z$ propagator, which is small with respect to $A$ anywhere but around the $Z$ resonance. Moreover, no contribution from SMEFT squared is present, except in the interference between $CP$ conserving and $CP$ violating dipole moments, and thus $C_{nk}$ is a perfect candidate to measure the sign and phase of the Wilson coefficients.  Only the real part of the Wilson coefficients contributes in the high-energy limit and indeed the lines for pure imaginary coupling deviate negligibly from the SM except around the $Z$ pole. It should be noted that, at high energy, information about the sign of the Wilson coefficient is lost, partially explaining the similar phenomenon in the quantum information observables. However, unlike the quantum observables, sizable deviations from the SM can be observed at lower energies, as well as separation between the lines belonging to different sign. It should be noted that, if we considered the $(nr)$ element instead, the roles of the imaginary and real part of the Wilson coefficient would be switched. This makes it straightforward to experimentally disentangle the phase of the Wilson coefficient, as real and imaginary part contribute to different entries of the spin correlation matrix. On the other hand, the quantum observables, being a combination of all the spin observables, lose differentiating power between the different scenarios for the phase of the Wilson coefficient.
\subsubsection{SMEFT Sensitivity}
\begin{figure}[htpb]
	\begin{subfigure}{1.1\textwidth}
	\centering
	\includegraphics[width=\textwidth]{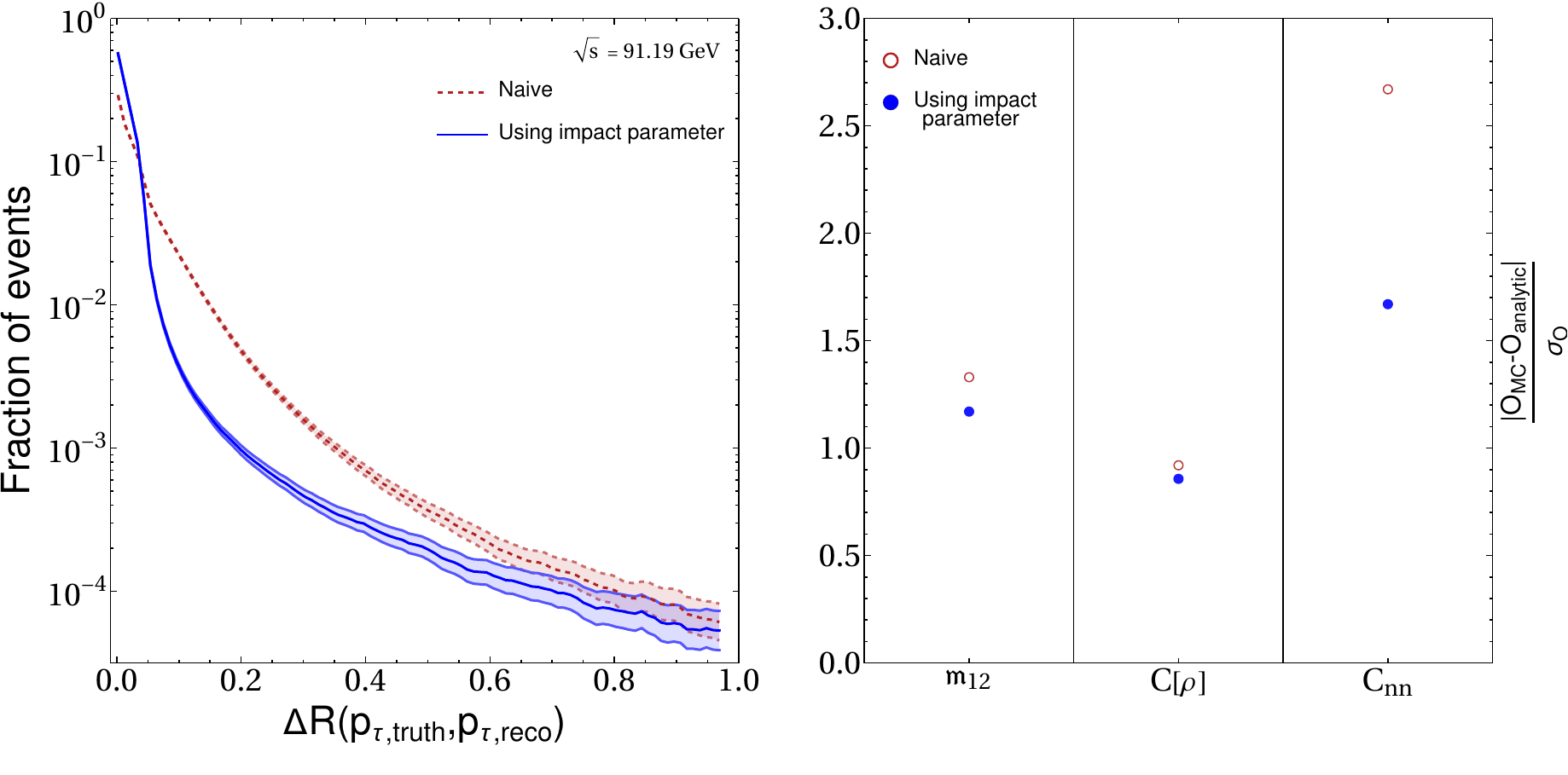}
	\end{subfigure}
	\begin{subfigure}{0.8\textwidth}
	\includegraphics[width=\textwidth]{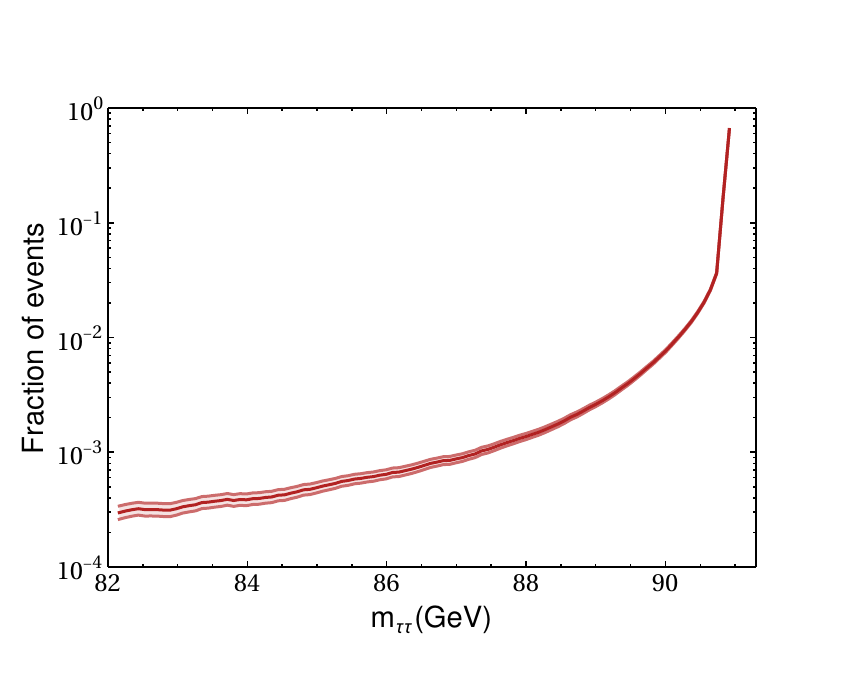}
	\end{subfigure}
	\caption{\textbf{Top Left}: Distribution of the angular separation $\Delta R$ between the true and reconstructed $\tau$ momentum, using a naive reconstruction, which is subject to a two-fold ambiguity (red,dashed), and taking into the account the information from impact parameters to resolve the ambiguity (blue, continuous). The bands show the statistical error of each bin.
		\textbf{Top Right}: deviation, in units of the statistical error, of the observables considered in this paper.
\textbf{Bottom}: $m_{\tau\tau}$ distribution from MC simulation including ISR effects, at a nominal beam energy of $\sqrt s = 91.19$ GeV. The band shows the statistical error in each bin.}
\label{fig:MomDistr}
\end{figure}

To compare the different observables as NP probes, we perform MC simulations using MadGraph~\cite{Alwall:2014hca}. We consider hadronic decays of the tau that only have one unobserved neutrino in the final state. We will consider only $\tau^\pm \to \pi^\pm \nu_\tau(\bar\nu_\tau)$ to obtain the sensitivity, but in \cref{sec:DetailsMC} we also comment on the three-body decay. We use the package TauDecay~\cite{Hagiwara:2012vz} for the decay form factors. The simulations are performed at three benchmark energies, namely $m_{\ell\ell}=10.56\;\textrm{GeV},m_Z, 500\; \textrm{GeV}$; the first energy is the $\Upsilon(4S)$ resonance at which Belle II operates, while the other two are proposed running energies for a TeraZ machine and one of the proposed future lepton colliders, respectively.
We find good agreement with the analytic results, which are at the level of undecayed taus, if ISR and detector effects are not taken into account. However, both ISR and detector effects introduce systematic shifts in the central values of the Fano coefficients. These shifts, while not large, can quickly dominate the error budget, and they are due to mis-reconstruction of the tau momenta, which, as shown in \cref{fig:MomDistr}, is improved by using impact parameters for the reconstruction~\cite{Jeans:2015vaa,Altakach:2022ywa}, see \cref{sec:ImpactPar}. ISR also impacts reconstruction by changing the di-tau mass energy, which does not coincide with the nominal center of mass energy anymore. The $m_{\tau\tau}$ distribution, calculated using Madgraph, is shown in \cref{fig:MomDistr}. We show in \cref{sec:DetailsMC} that integrating over this distribution generates a small shift of order $\lesssim 1$\textperthousand.
Details and results of the MC simulations are collected in \cref{sec:DetailsMC}.

\begin{figure}[t]
	\centering
	\includegraphics[width=\textwidth]{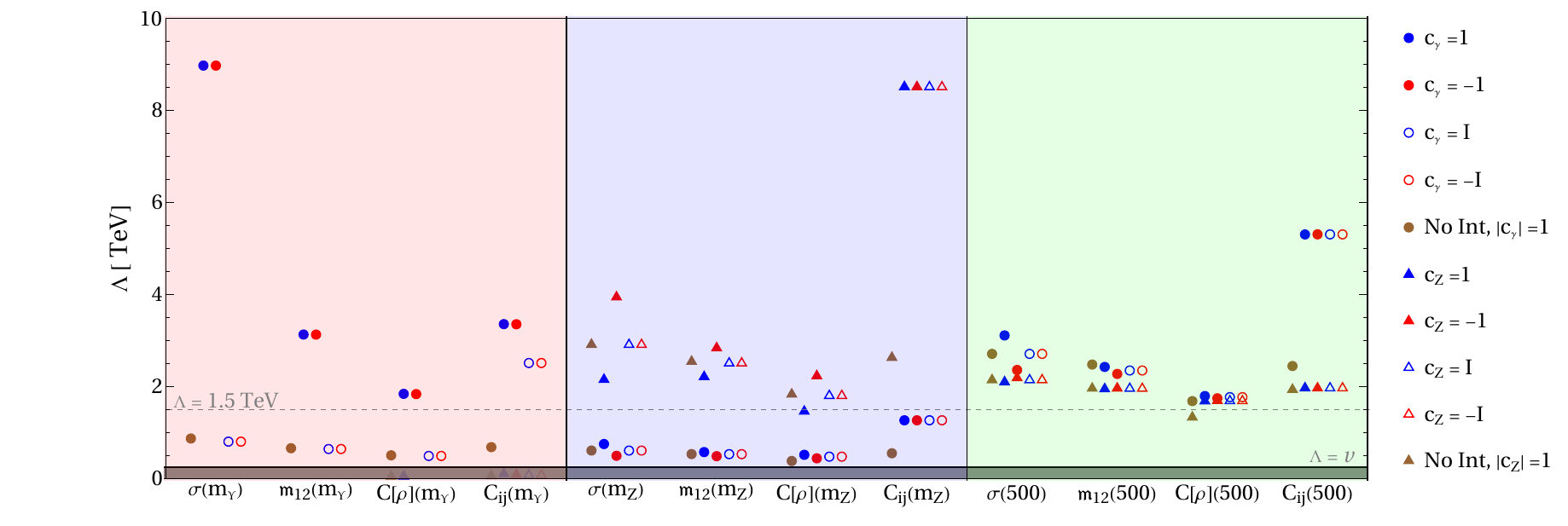}
	\caption{95\% CL sensitivity on the new physics scale $\Lambda = c_{\gamma/Z}/\sqrt{C_{\gamma/Z}} $ for the $\mathcal{O}_{\gamma/Z\, \tau }$ operators, derived from measurements of the total cross section, as well as different spin correlation observables at three benchmark energies. Quantum information observables perform worse than the single coefficient of the $C$ matrix at all energies. At large energies, the $C$ matrix affords a greater sensitivity than the other observables by a factor 2-3, because of the resurrection phenomenon, see main text. NP scales lower than the EW scale (shaded gray region) break the EFT assumption.
	}
	\label{fig:plLambdas}
\end{figure}

We use the statistical uncertainties estimated from MC simulations to derive the projected sensitivities to dipole operators, shown in \cref{fig:plLambdas}. At all benchmark energies, we also show the reach of the total cross section $\sigma$. Realistic projections should include an estimate of the systematic uncertainties, to be computed using a full detector simulation, which is beyond the scope of this paper. Indeed, we do not expect our conclusions regarding the relative power of the different observables to be impacted by systematic uncertainties, although the experimental reach in \cref{fig:plLambdas} might prove to be too optimistic when systematic effects are taken into account.

It is clear from \cref{fig:plLambdas} that, at energies high enough to be sensitive to the CP violation in the EW theory, the quantum information observables diminish the sensitivity with no advantage. At Belle II energies, the concurrence has a slightly better sensitivity for real couplings compared to the spin correlations, but at the price of significantly worse sensitivity for imaginary couplings. Moreover, the bounds are derived from one single element of the $C$ matrix; the sensitivity can be enhanced by considering more than one entry, as well as the polarisation vectors $B$. Moreover, the Fano coefficient can be used to construct observables that are optimal in the sense of having the minimal possible statistical uncertainties~\cite{Atwood:1991ka,Davier:1992nw,Diehl:1993br}, see~\cite{Bernreuther:2021elu} for a recent analysis of the tau dipole operators using optimal observables\footnote{We thank an anonymous referee for introducing the author to the concept of optimal observables.}.

Thus, we argue that the quantum observables perform worse in resurrections scenarios, and stronger bounds are instead provided by the spin correlation matrix directly.

Quantum information observables do not display increased sensitivity to the phase of the Wilson coefficient, and are unable to distinguish CP conserving and violating scenarios. Due to their definition, which involves taking the eigenvalues of a $3\times 3$ matrix, it is challenging to undertsand the origin of this facts. The $C$ matrix coefficients, instead, can be easily calculated and interpreted. Experimentally, measuring the $C$ matrix is needed to reconstruct the density matrix, so there seems to be no advantage in using quantum information observables in this scenario.

Using \cref{eq:AnMoms} we can translate the SMEFT sensitivity into sensitivity to anomalous magnetic and electric dipole moments of the $\tau$. The results are collected in \cref{tab:AnMoms}. We can see that, using only one Fano coefficient, we obtain sensitivity that is roughly a factor of 3 worse than the analytical predictions, obtained in~\cite{Bernreuther:2021elu}, for optimal observables at Belle II using the same decay channell\@. However, it can be seen that an $e^+ e^-$ collider running at $500$ GeV has a sensitivity that is roughly a factor of 2 better than the one estimated for a $10$ TeV muon collider~\cite{Denizli:2024uwv}. Finally, the sensitivity on the weak electromagnetic dipole moment exceed by more than an order of magnitude those computed in~\cite{Fabbrichesi:2025ywl} for LEP3 running at $\sqrt{s} = 240$ GeV using the trace distance between states.

\begin{table}
	\centering
$\begin{array}{|c|c|c|c|c|}
	\hline
	& \Delta a_{\tau} (\times 10^{-4}) & \left|\Delta d_{\tau}\right| (\times 10^{-18}\, e\, \textrm{cm}) &
	 \Delta a_{\tau}^Z (\times 10^{-4}) & \left|\Delta d_{\tau}^Z\right| (\times 10^{-18}\, e\, \textrm{cm}) \\
	\hline
	\sqrt{s} = m_{\Upsilon} & 0.50 & 3.6 & 350 & 396\\
	\hline
	\sqrt{s} = m_{Z} & 25 & 14 & 0.55 & 0.31\\
	\hline
	\sqrt{s} = 500\,\textrm{GeV}  & 1.4 & 0.81 & 10 & 5.8\\
	\hline
\end{array}$
\caption{Sensitivity to anomalous dipole moments of the $\tau$ using the most sensitive observable out of those we consider in this paper (see \cref{fig:plLambdas}), at three benchmark energies. The obtained bounds on the weak dipole moments at $\sqrt{s} = m_{\Upsilon}$ break the EFT assumption and as such should not be considered as reliable.}
	\label{tab:AnMoms}
\end{table}
\subsubsection{Hadron Colliders}
\label{sec:Hadrons}
\begin{figure}[t]
	\centering
	\includegraphics[width=1.2\textwidth]{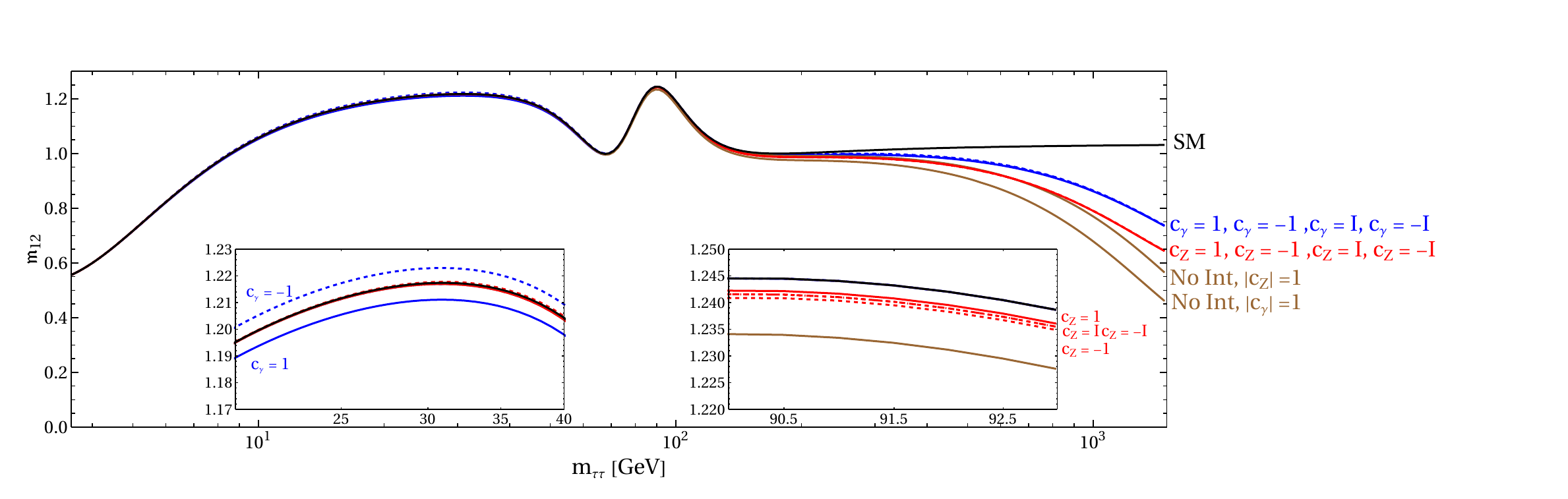}
	\caption{Same as \cref{fig:m12e}, but for $pp$ initial state.}
	\label{fig:m12p}
\end{figure}

\begin{figure}[t]
	\centering
	\includegraphics[width=1.2\textwidth]{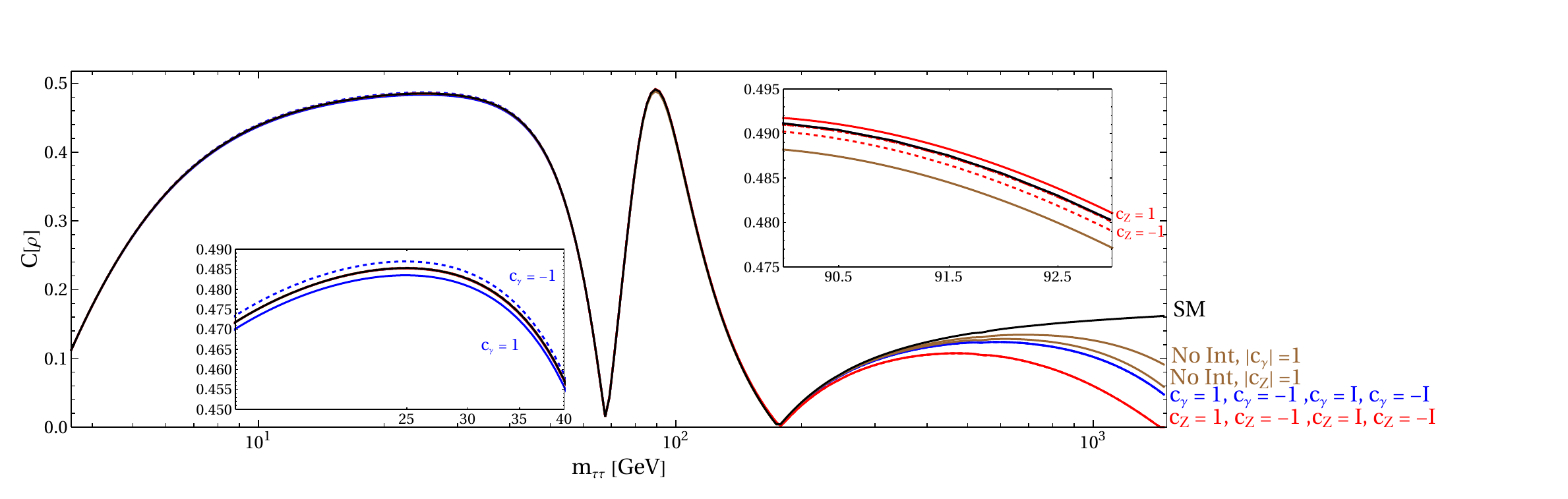}
	\caption{Same as \cref{fig:conce}, but for $pp$ initial state.}
	\label{fig:concp}
\end{figure}

\begin{figure}[t]
	\centering
	\includegraphics[width=1.2\textwidth]{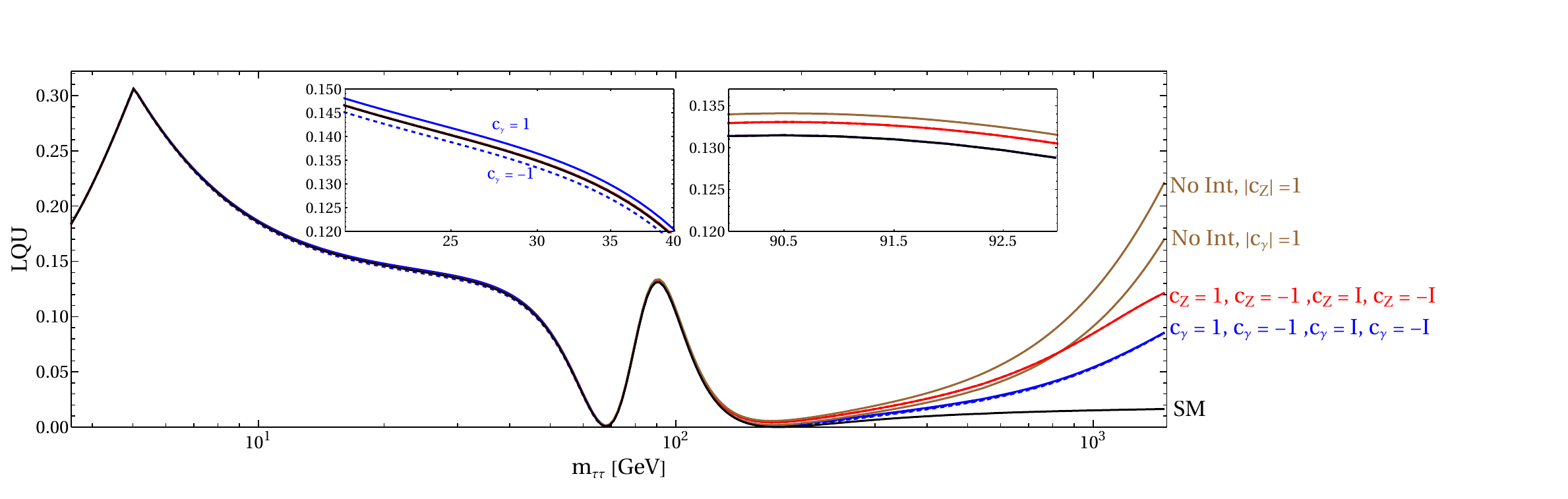}
	\caption{Same as \cref{fig:lque}, but for $pp$ initial state.}
	\label{fig:lqup}
\end{figure}

\begin{figure}[t]
	\centering
	\includegraphics[width=1.2\textwidth]{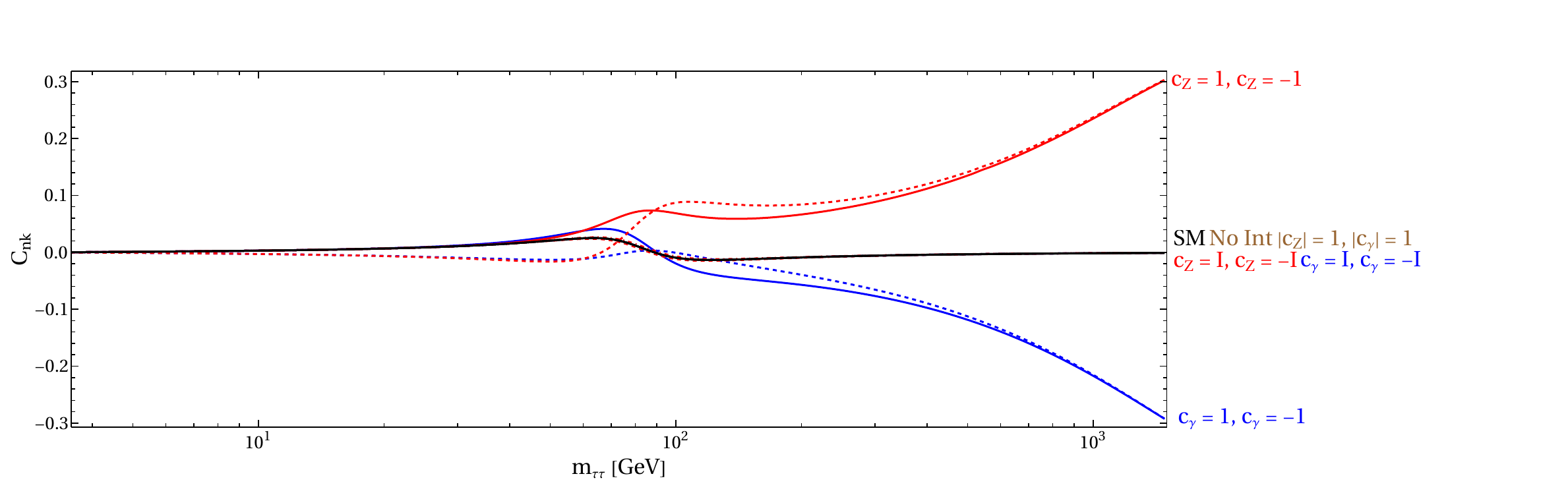}
	\caption{Same as \cref{fig:c23e}, but for $pp$ initial state.}
	\label{fig:c23p}
\end{figure}

At hadron colliders, the LO prediction at partonic level has to be weighted by the PDF of the partons, as explained in \cref{sec:formalism}. \Cref{fig:m12p,fig:concp,fig:lqup,fig:c23p} are the analogue of \cref{fig:m12e,fig:conce,fig:lque,fig:c23e}, and the same patterns can be seen regarding resurrection of interference. Namely, the net effect of the resurrection phenomenon in the Bell inequality marker $\mathfrak{m}_{12}$ and the local quantum uncertainty LQU is to decrease the sensitivity to the NP scale $\Lambda$. For the concurrence $C[\rho]$ the resurrection phenomenon improves the sensitivity, but only for real Wilson coefficients. In the spin correlation matrix elements, instead resurrection improves the sensitivity, both for real and imaginary Wilson coefficients.

As mentioned in \cref{sec:Tomography}, the measurement of spin correlations at hadron colliders is much more involved than at lepton colliders. While at lepton colliders the ambiguity due to the non-measurement of neutrinos can be resolved using impact parameters~\cite{Jeans:2015vaa}, the same task is made more difficult at hadron colliders by the lack of knowledge of the c.o.m.~frame. Measurements at  resonances, such as $m_{\ell \ell} = m_Z (m_H)$ can partially circumvent this difficulty by assuming the process predominantly occurs via on-shell $Z(H)$ decay~\cite{Elagin:2010aw}. Indeed, a measurement of the $\tau$ polarisation has recently been performed at the CMS experiment, using a sample with an integrated luminosity of 36.3 $fb^{-1}$~\cite{CMS:2023cqh}. Different strategies have been devised to boost the reconstruction efficiency for the $\tau$ momentum, including chi-squared methods~\cite{Chen:2018cxt}, matrix element~\cite{Bianchini:2016yrt} and machine learning~\cite{Bartschi:2019xlg,Tamir:2023aiz,Tani:2024qzm} techniques, and a dedicated algorithm to calculate the spin correlation coefficients at hadron collider exists~\cite{Korchin:2024uxy}. 

Given the difficulties posed by a measurement of polarisations at collider, we do not try here to estimate the experimental uncertainty on the spin correlation coefficients. Instead, we report the uncertainty that a Z-pole measurement should have to reach the benchmark bound $C_{\gamma} \lesssim 1/1.5\, \mathrm{TeV}^2$ (corresponding to $\Delta a_{\tau} < 1.78 \times 10^{-3}$), obtained from high-energy measurement of the cross section. That is, to be competitive with the limit reported in Ref.~\cite{Haisch:2023upo}  at 95\%CL, the uncertainty $\Delta C (\Delta B)$ on the entries of $C (B)$ needs to satisfy
\begin{align}
	\Delta B < 0.0005 \, ,
	\label{eq:Senspp} \\
	\Delta C < 0.008\, ,
\end{align}
to be contrasted with the current uncertainty of $\Delta B_{\textrm{CMS}} = 0.015$~\cite{CMS:2023cqh}. Similarly to the $ee$ case, the phase of the Wilson coefficient can be determined by measuring all the entries of $B$ and $C$. For the weak dipole, an exclusion of $C_Z \lesssim 1/1.5\, \mathrm{TeV}^{-2}$ is expected if the following experimental uncertainties can be reached
\begin{align}
	\Delta B < 0.003 \label{eq:Senspp2}\, ,\\
	\Delta C < 0.03\,.
\end{align}

In general, as it can be seen from the Fano coefficients in \cref{sec:Fano}, the constraint on $C_{\gamma/Z}$ scales linearly with $\Delta B$ and $\Delta C$, so the constraint on the SMEFT scale $\Lambda$ scales with the square root of the experimental precision, which in turn scales with the square root of the number of events, making it challenging to increase the sensitivity on $\Lambda$, even if systematic effects are under control.

%%%%%%
\section{Conclusions and Outlook}
\label{sec:conclusions}
In this paper we explored interference resurrection in 4-fermion scattering by measuring the final-state angular correlations, and the resulting improvement in sensitivity to SMEFT operators.
We find that, at moderate and large energies compared to the EW scale, spin observables greatly outperform the cross section in sensitivity to the new physics scale, improving it by up to a factor of 2-3.
We also consider quantum information observables, which have obtained considerable attention lately in the top sector, and have been shown to outperform other angular observables as a probe of SMEFT operators~\cite{Maltoni:2024tul}. 
We compare the sensitivity on the new physics scale $\Lambda$ for $\gamma$ or $Z$ dipole operators, in \cref{fig:plLambdas}, and find that there is no advantage in employing quantum observables like the concurrence $C[\rho]$, the Bell inequality violation marker $\mathfrak{m}_{12}$ or the local quantum uncertainty LQU\@. 
Indeed, simple spin correlations significantly improve the sensitivity in comparison to both cross-section measurement and quantum information observables.
As we show in \cref{fig:m12e,fig:conce,fig:lque,fig:c23e} for the $ee$ case and \cref{fig:m12p,fig:concp,fig:lqup,fig:c23p} for the $pp$ case, the resurrection is only partially successful in the quantum information observables, because phase information is lost, and in some cases the interference partially cancels against the SMEFT squared contribution.

We believe that the present results are at odds with those regarding the top pair case because top pair production respects $CP$, $C$ and $P$ parities to a good approximation~\cite{Afik:2020onf}, which imposes a quite simple spin state.
In this case, entanglement can be directly extracted from the diagonal entries of the spin correlation matrix $C$.
However, as we observed in~\cref{sec:ToyModel}, resurrection of interference only involves the off-diagonal entries of the $C$ matrix, so the simplified entanglement markers do no participate in resurrection.
Moreover, the violation of $P$ and $C$ symmetries makes the simplified criterion for entanglement used in~\cite{Maltoni:2024tul} only a sufficient, but not necessary, condition for entanglement.
More appropriate entanglement markers are $\mathfrak{m}_{12}$ and $C[\rho]$, that distill the $R$ matrix to a single number, thus losing the ability to differentiate the different off-diagonal components of the $C$ matrix.
In addition, the relatively complicated expressions of $\mathfrak{m}_{12}$ and $C[\rho]$, involving the eigenvalues of a $3\times 3$ matrix, hinder an analytical understanding of how exactly the quantum information observables relate to the underlying spin coefficients.
We can nonetheless observe in \cref{fig:m12e,fig:conce,fig:lque,fig:c23e,fig:m12p,fig:concp,fig:lqup,fig:c23p} that $CP$ information is mostly lost in the quantum information observables at high energies, while the spin correlation coefficients retain it.
This support the hypothesis that the apparent superiority of quantum information observables in top pair production is due to the absence of $CP$ violation. In that case, a simple combination of the $C$ matrix elements, the opening angle coefficient $D$ (see \cref{eq:simplified}), is at the same time an entanglement marker and a good probe of NP.
On the contrary, in the case of EW tau pair production, the quantum information observables are a more complicated combination of the spin correlation coefficients, and they end up being suboptimal probes for $CP$ violation and NP\@.
Since the spin correlation coefficients are anyway the building blocks for any quantum information study, they seem like a much more appropriate choice as NP probes.\footnote{ Ref.~\cite{Sullivan:2024wzl}, which appeared later than the present work, presents similar finding for $h-> V V.$}

Looking forward, the identification of optimal observables that include systematic effects for the set of operators discussed in this work is an important endeavor.
Indeed, it has been shown~\cite{Chai:2024zyl} that the statistically optimal observables of Ref.~\cite{Atwood:1991ka,Diehl:1993br,Davier:1992nw} introduce biases when systematic effects and background are taken into account.
For this goal, realistic estimations of the sensitivity of current and proposed colliders to spin correlation is necessary.
While studies for lepton collider are abundant, fewer exist for hadron collider, due to the known difficulty of correctly reconstructing the final particles' 4-momentum when invisible particles are present in  their decay.
Recent efforts~\cite{Bianchini:2016yrt,Bartschi:2019xlg,Tamir:2023aiz,Tani:2024qzm,Korchin:2024uxy} might change the situation, but the needed sensitivities to be competitive with cross-section measurements seem at this time far in the future, see the discussion in \cref{sec:Hadrons}.

All the calculations performed in this work are at tree-level, given that the effect of NLO is expected to be small, as spin correlations are ratios of cross sections.
Nonetheless, it would be interesting to understand the effect of soft radiation and loop effects. Interestingly, one expects the properly resummed tree-level answer, corresponding a classical field theory, not to exhibit entanglement at all.
Thus, if the LO calculation is to be close to the exact one, it seems like quantum corrections, on their own, would work to increase entanglement.
The interplay of quantum corrections and soft radiation merits further discussion.

Entanglement has also been proposed as a guiding principle for parameter selection through the emergence of global symmetries via a sort of stationary principle for entanglement~\cite{Cervera-Lierta:2017tdt,Beane:2018oxh,Low:2021ufv,Liu:2022grf,Carena:2023vjc,Maltoni:2024csn,Kowalska:2024kbs}.
While enticing, most of these proposals either rely on specific angular configurations or additional assumptions (see the discussion in Ref.~\cite{Kowalska:2024kbs}), so at the current state it is hard to establish a general principle and more information is needed.
In this paper, we add another piece to the puzzle by studying the LQU, a discord-like measure for non-classical correlation~\cite{Girolami:2012tz}.
As can be seen in~\cref{fig:lque,fig:lqup}, the LQU has a local maximum that is absent in the entanglement markers, and cannot be easily explained by the presence of a symmetry or resonance.
While other measures of quantum discord have already been considered in the literature~\cite{Afik:2022dgh}, their simplified definition used so far can only be applied to the simple spin state of the top, and the complete case involves a complicated minimisation procedure.
In contrast, the LQU has an explicit formula that is applicable to any bipartite qubit~\cite{Girolami:2012tz}, so its use facilitates studies of non-classical correlations beyond entanglement.
It will be interesting to study the robustness of the principles proposed in~\cite{Cervera-Lierta:2017tdt,Beane:2018oxh,Low:2021ufv} on measures of non-classicality that go beyond entanglement using the LQU\@. 

\begin{acknowledgments}
The author is very grateful to Felix Yu for many useful discussions.  
The author acknowledges a conversation with Miki Chala about resurrection.
The author would like to thank Felix Yu and Sebastian Schenk for feedback on the manuscript.
The author would also like to thank the Fermilab theory group for their kind hospitality while this work was in progress.
The author is grateful to an anonymous referee for comments on a previous version of the manuscript that helped significantly improve this work. 
This work was supported by the Cluster of Excellence {\em Precision Physics, Fundamental Interactions and Structure of Matter} (PRISMA${}^+$ -- EXC~2118/1) within the German Excellence Strategy (project ID 390831469).
\end{acknowledgments}
\appendix

\section{Fano Coefficients}
	\label{sec:Fano}
	Explicit expression for the $R$ matrix in the electroweak SM, as well as with a parametrisation of the effect of anomalous dipole contributions, first appeared in~\cite{Bernreuther:1992be}, and we agree with their results\footnote{We thank the anonymous referee for informing us of this paper.}. More recently, they appeared~\cite{Banerjee:2022sgf}, but they were not calculated in the helicity basis that we adopt here. 
	Finally, explicit expressions for the Fano coefficients in the helicity basis have also appeared in~\cite{Maltoni:2024tul}; we agree with their results on most terms, but we find that some non-vanishing off-diagonal correlation that were not considered in~\cite{Maltoni:2024tul}. These terms are due to the non-zero $Z$ width, and they generate absorptive contributions to the matrix elements, which cannot be generated from the $\Gamma_Z = 0$ expression by the usual substitution $m_Z^2 \to m_Z^2 - i m_Z\Gamma_Z$. Of course, contributions linear in $\Gamma_Z$ are formally one loop and small in the SM due to the small ratio $\Gamma_Z/M_Z$, but nonetheless they modify the spin configuration of the final state, and might be relevant for other vector mediators with larger widths. Moreover, we explicitly write down the polarisation vector $B$, which has not appeared elsewhere, although it can be obtained from the results of~\cite{Bernreuther:1992be}.

	Following~\cite{Maltoni:2024tul} we divide the contributions according to their tensor structure, that is we split the spin production matrix in three parts:

	\begin{equation}
		R = R^{[0]} + R^{[1]} + R^{[2]} \, ,
		\label{eq:RGammaIns}
	\end{equation}
	where the superscript indicates the number of $\gamma^5$ insertions on the $\tau$ line. This is because the tensor structure uniquely defines the spin state. Moreover, still following~\cite{Maltoni:2024tul}, we denote common factors by $F$. Throughout this section, the coefficients that are not explicitly written are either zero or can be obtained through the known symmetry of the Fano coefficients~\cite{Bernreuther:2015yna}.

	We start with the SM\@. At zero $\gamma^5$ insertions, the Fano coefficients are:
	\begin{equation}
		\begin{cases}
			A^{[0]} &= F^{[0]}(\beta_\tau^2 z^2 - \beta_\tau^2 + 2)\\
			\tilde{B}^{[0]}_r &= F_B^{[0]}\sqrt{1-z^2}\sqrt{1-\beta_\tau^2} \\
			\tilde{B}^{[0]}_k &= -F_B^{[0]}z \\
			\tilde{C}^{[0]}_{nn} &= -F^{[0]}\beta_\tau^2 (1-z^2)\\
			\tilde{C}^{[0]}_{rr} &= -F^{[0]}(\beta_\tau^2-2)(1-z^2)\\
			\tilde{C}^{[0]}_{kk} &= F^{[0]}(\beta_\tau^2 -(\beta_\tau^2-2)z^2)\\
			\tilde{C}^{[0]}_{kr} &= 2F^{[0]}\sqrt{1-\beta_\tau^2}z\sqrt{ 1-z^2 }\\
		\end{cases}\, , 
		\label{eq:CoeffSM0}
	\end{equation}
	where the common factors are
	\begin{align}
		F^{[0]} &=  4 N e^4\left( Q_\tau^2Q_f^2 + 8\Re \left(\frac{ Q_\tau Q_f Q_{V\tau}Q_{Vf} m_\tau^2}{D_Z}\right) + \frac{16 Q_{V\tau}^2(Q_{Vf}^2+Q_{Af}^2)m_\tau^4}{|D_Z|^2} \right)\, , \\
		F^{[0]}_B &=  64 N e^4\left(\Re\left(\frac{Q_{Af}Q_{V\tau}Q_fQ_\tau m_\tau^2}{D_Z}\right) + 4 \frac{Q_{Af}Q_{Vf}Q_{V\tau}^2 m_\tau^4}{|D_Z|^2} \right) \, ,
		\label{eq:FSM0}
	\end{align}
	with $N$ the normalisation factor in \cref{eq:R}, and $D_Z = c_Ws_W\left( 4m_\tau^2 - (1-\beta_\tau^2)(m_Z^2- im_Z\Gamma_Z) \right)$ the boosted propagator of the $Z$ boson, together with the electroweak mixing angles coming from the coupling to fermions. 

	At one insertion of $\gamma^5$, we see the first absorptive contributions, noted by the $\Gamma$ subscript. They do not share the same common factor with the rest of the coefficients, and disappear as $\Gamma \to 0$:

	\begin{equation}
		\begin{cases}
			A^{[1]} &= 2F^{[1]}z\\
			\tilde{B}^{[1]}_n &= F_{\Gamma,B}^{[1]}\sqrt{1-\beta_\tau^2}\sqrt{1-z^2} \\
			\tilde{B}^{[1]}_r &= F_B^{[1]}\sqrt{1-\beta_\tau^2}z\sqrt{ 1+z^2 }\\
			\tilde{B}^{[1]}_k &= F_B^{[1]}(1+z^2) \\
			\tilde{C}^{[1]}_{kk} &= 2F^{[1]}z\\
			\tilde{C}^{[1]}_{nr} &= F^{[1]}_\Gamma (1-z^2)\\
			\tilde{C}^{[1]}_{nk} &= -F_\Gamma^{[1]}\sqrt{1-\beta_\tau^2}z\sqrt{ 1-z^2 }\\
			\tilde{C}^{[1]}_{rk} &= F^{[1]}\sqrt{1-\beta_\tau^2}\sqrt{ 1-z^2}\\
		\end{cases}\, ,
		\label{eq:CoeffSM1}
	\end{equation}
	\begin{align}
		F^{[1]} &=  16 N Q_{A\tau}Q_{Af}Q_{A\tau}\beta_\tau e^4\left( 2\Re \left(\frac{Q_\tau Q_f m_\tau^2}{D_Z}\right) + \frac{16 Q_{V\tau}^2Q_{Vf}m_\tau^4}{|D_Z|^2}  \right)\, , \\
		F^{[1]}_B &=  32 N e^4\beta_\tau Q_{A\tau}\left( \Re\left(\frac{Q_{Vf}Q_fQ_\tau m_{\tau}^2 }{D_Z} \right) + 4 \frac{Q_{V\tau}(Q_{Vf}^2+Q_{Af}^2)m_\tau^4}{|D_Z|^2} \right) \, , \\
		F^{[1]}_\Gamma &=  32 N e^4\beta_\tau\,\Im\left(\frac{Q_{A\tau}Q_{V\tau}Q_fQ_\tau c_W^2 s_W^2 m_\tau^2}{D_Z}\right) \, ,\\
		F^{[1]}_{\Gamma,B} &=  F^{[1]}_\Gamma \; \textrm{with} \; \left( Q_{Vf} \to Q_{Af} \right)  \, .
		\label{eq:FSM1}
	\end{align}

	The contributions from double $\gamma^5$ insertions are:
	\begin{equation}
		\begin{cases}
			A^{[2]} &= F^{[2]}(1+z^2)\\
			\tilde B^{[2]}_k &= F_B^{[2]}z\\
			\tilde C^{[2]}_{nn} &= F^{[2]}(1-z^2)\\
			\tilde C^{[2]}_{rr} &= -F^{[2]}(1-z^2)\\
			\tilde C^{[2]}_{kk} &= F^{[2]}(1+z^2)\\
		\end{cases}\, ,
		\label{eq:CoeffSM2}
	\end{equation}
	\begin{align}
		F^{[2]} &=  64 N e^4 Q_{A\tau}^2\beta_\tau^2 \frac{(Q_{Vf}^2+Q_{Af}^2)m_{\tau}^4}{|D_Z|^2}\, ,\\
		F^{[2]}_B &=  -256 N e^4\left(\frac{Q_{Af}Q_{Vf}Q_{A\tau}^2m_\tau^4}{|D_Z|^2}\right) \, .
		\label{eq:FSM2}
	\end{align}

	We now pass to the dipole contribution. Similarly to the SM, we find absorptive contributions that are linear in $\Gamma_Z$; we consider the general case of complex couplings.

	At one dipole insertion (\textit{i.e.}~one insertion of $\sigma^{\mu\nu}$), only zero or one $\gamma^5$ insertions are possible. At zero insertions we get

	\begin{equation}
		\begin{cases}
			A^{[6,0]} &= F^{[6,0]}\\
			\tilde{B}^{[6,0]}_r &= F_B^{[6,0]}\dfrac{\beta_\tau^2-2}{\sqrt{1-\beta_\tau^2}}\sqrt{1-z^2} - F^{[6,0]}_{\Gamma,B}\dfrac{z\sqrt{1-z^2}}{\sqrt{1-\beta_\tau^2}}\\
			\tilde{B}^{[6,0]}_n &= F_B^{[6,0]}\dfrac{\beta_\tau}{\sqrt{1-\beta_\tau^2}}\sqrt{1-z^2} - F^{[6,0]}_{\Gamma,B}\dfrac{z\sqrt{1-z^2}\beta_\tau}{\sqrt{1-\beta_\tau^2}}\; \textrm{with}\;\left( \Re(c_{\gamma/Z}) \leftrightarrow \Im(c_{\gamma/Z}) \right)\\
			\tilde{B}^{[6,0]}_k &= 2F_B^{[6,0]}z - F^{[6,0]}_{\Gamma,B}(1-z^2)\\
			\tilde{C}^{[6,0]}_{rr} &= F^{[6,0]}(1-z^2)\\
			\tilde{C}^{[6,0]}_{kk} &= F^{[6,0]}z^2\\
			\tilde{C}^{[6,0]}_{rn} &= F^{[6,0]}\dfrac{\beta_\tau}{2}(1-z^2)\; \textrm{with}\; \left(\Re (c_{\gamma/Z}) \to \Im(c_{\gamma/Z}) \right)\\
			\tilde{C}^{[6,0]}_{rk} &= F^{[6,0]}\dfrac{2- \beta_\tau^2}{2\sqrt{1-\beta_\tau^2}}z\sqrt{1-z^2}- F^{[6,0]}_\Gamma \dfrac{\sqrt{1-z^2}}{\sqrt{1-\beta_\tau^2}}\\
			\tilde{C}^{[6,0]}_{nk} &= F^{[6,0]}\dfrac{\beta_\tau^2\sqrt{1-z^2}}{\sqrt{1-\beta_\tau^2}} + F^{[6,0]}_\Gamma \dfrac{\beta_\tau z\sqrt{1-z^2}}{2\sqrt{1-\beta_\tau^2}}\; \textrm{with} \; \left( \Re(c_{\gamma/Z}) \leftrightarrow \Im(c_{\gamma/Z}) \right)\\
		\end{cases}\, ,
		\label{eq:Coeff60}
	\end{equation}

	\begin{align}
		F^{[6,0]} &=  16\sqrt2 N e^3 \frac{m_\tau v}{\Lambda^2}\biggr(\Re(c_{\gamma})Q_f^2Q_\tau - \Re(c_{Z})\frac{16 c_W s_W Q_{V\tau}(Q_{A f}^2 + Q_{Vf}^2) m_\tau^4}{{|D_Z|}^2}
		\label{eq:F60}\\
	\nonumber	  &+ \left(\Re(c_{\gamma})Q_{V \tau} - \Re(c_{Z})Q_\tau c_W s_W \right)\Re\left(\dfrac{4 Q_{V f}Q_{f}m_\tau^2}{D_Z}\right) \biggr) \, ,\\
	F^{[6,0]}_\Gamma &=  8\sqrt2 N e^3\beta_\tau \frac{m_\tau v}{\Lambda^2}
	\Im\left(c_{\gamma}Q_{V \tau} + c_{Z}Q_\tau c_W s_W\right) \Im\left(\dfrac{Q_{A f}Q_{f}m_\tau^2}{D_Z}\right) \, ,\\
	\label{eq:FGamma60}
	F^{[6,0]}_B &= 8\sqrt2 N e^3\beta_\tau \frac{m_\tau v}{\Lambda^2}  \biggr( Q_{Af}Q_{f}\left(  \Re\left(c_{\gamma}\right)Q_{V \tau} + \Re\left(c_{Z}\right)Q_\tau c_W s_W \right)\Re\left(\frac{m_\tau^2}{D_Z}  \right) \\
\nonumber  & + 8Q_{Af}Q_{Vf}Q_{V\tau}\Re\left(c_{Z}\right)\frac{m_\tau^4}{|D_Z|^2}  \biggr) \, ,\\
F^{[6,0]}_{\Gamma,B} &= F^{[6,0]}_{\Gamma}\; \textrm{with} \; \left( Q_{Af} \to Q_{Vf} \right)\,  .
\end{align}

At one insertion of both $\gamma^5$ and $\sigma^{\mu\nu}$, we obtain:
\begin{equation}
	\begin{cases}
		A^{[6,1]} &= F^{[6,1]}z\\
		\tilde{B}^{[6,1]}_r &= F^{[6,1]}_{B}\dfrac{z\sqrt{1-z^2}}{\sqrt{1-\beta_\tau^2}} + F^{[6,1]}_{\Gamma,B}\dfrac{\beta_\tau}{\sqrt{1-\beta_\tau^2}}\sqrt{1-z^2} \\
		\tilde{B}^{[6,1]}_r &= F^{[6,1]}_{B}\dfrac{z\sqrt{1-z^2}}{\sqrt{1-\beta_\tau^2}} + F^{[6,1]}_{\Gamma,B}\dfrac{1}{\sqrt{1-\beta_\tau^2}}\sqrt{1-z^2} \; \textrm{with} \; \left( \Re(c_{\gamma}) \leftrightarrow \Im(c_{\gamma}) \right) \\
		\tilde{B}^{[6,1]}_k &= F^{[6,1]}_{B} (1+z^2)\\ 
		\tilde{C}^{[6,1]}_{kk} &= F^{[6,1]}z\\
		\tilde{C}^{[6,1]}_{nr} &= F^{[6,1]}_{\Gamma} (1-z^2)\\
		\tilde{C}^{[6,1]}_{rk} &= F^{[6,1]}\dfrac{\sqrt{1-z^2}}{2\sqrt{1-\beta_\tau^2}} + F^{[6,1]}_\Gamma \dfrac{1}{\sqrt{1-\beta_\tau^2}} z\sqrt{1-z^2}\\
		\tilde{C}^{[6,1]}_{nk} &= F^{[6,1]}\dfrac{\beta_\tau\sqrt{1-z^2}}{2\sqrt{1-\beta_\tau^2}} -  F^{[6,1]}_\Gamma\dfrac{\beta_\tau}{\sqrt{1-\beta_\tau^2}} z\sqrt{1-z^2} \; \textrm{with}\; \left( \Re(c_{\gamma/Z}) \to \Im(c_{\gamma/Z}) \right)\\
	\end{cases}\, ,
	\label{eq:Coeff61}
\end{equation}

\begin{align}
	F^{[6,1]} &=  64\sqrt2 N e^3 (\frac{m_\tau v}{\Lambda^2})Q_{A\tau}Q_{A f}\beta_\tau m_\tau^2 \left(\Re(c_{\gamma})\Re(\frac{Q_f m_{\tau}^2}{D_Z}) - \Re(c_{Z})\frac{8 c_W s_W Q_{V\tau} m_\tau^4}{{|D_Z|}^2} \right. \label{eq:F61}\\
\nonumber	   &+ \left(\Re(c_{\gamma})Q_{V \tau} - \Re(c_{Z})Q_\tau c_W s_W \right)\Re\left(\frac{4 Q_{V f}Q_{f}m_\tau^2}{D_Z}\right) \biggr)\, ,\\
F^{[6,1]}_\Gamma &=  32\sqrt2 N e^3 (\frac{m_\tau v}{\Lambda^2}) Q_{Vf}Q_{A\tau}Q_{f}\beta_\tau \Re(c_{\gamma} c_W^2 s_W^2)\Re(\frac{m_\tau^2}{D_Z})\, ,\\
F^{[6,1]}_B &=  32\sqrt2 N e^3 Q_{A\tau}(\frac{m_\tau v}{\Lambda^2}) \beta_\tau \left(\Re(c_{\gamma}Q_{f}Q_{Vf})\Re(\frac{m_{\tau}^2}{D_Z}) + \Re(c_{Z})\frac{4 c_W s_W (Q_{Vf}^2+Q_{A f}^2) m_\tau^4}{{|D_Z|}^2} \right) \, ,\\
F^{[6,1]}_{\Gamma,B} &= F^{[6,1]}_{\Gamma}\; \textrm{with} \; \left( Q_{Af} \to Q_{Vf}\, , \Re(c_{\gamma}) \to \Im(c_{\gamma}) \right)\,  .
\end{align}

For the double insertion of the dipole operators, we find it useful to split contributions by counting the numbers of factors of $\Im(c_{\gamma/Z})$, keeping the same notation above. Indeed, insertion of the imaginary part of the Wilson coefficient are essentially a $\gamma^5$ insertion. At zero insertions we obtain:
\begin{equation}
	\begin{cases}
		A^{[8,0]} &= F^{[8,0]}(-\beta_\tau^2z^2 - \beta_\tau^2 + 2)\\
		\tilde{B}^{[8,0]}_r &= F^{[8,0]}_{B} \dfrac{\sqrt{1-z^2}}{\sqrt{1-\beta_\tau^2}}\\
		\tilde{B}^{[8,0]}_k &= F^{[8,0]}_{B} z \\
		\tilde{C}^{[8,0]}_{rr} &= F^{[8,0]}(2-\beta_\tau^2)(1-z^2)\\
		\tilde{C}^{[8,0]}_{nn} &= F^{[8,0]}\beta_\tau^2(1-z^2)\\
		\tilde{C}^{[8,0]}_{kk} &= F^{[8,0]}\left( (2-\beta_\tau^2)z^2 - \beta_\tau^2 \right)\\
		\tilde{C}^{[8,0]}_{rk} &= F^{[8,0]}2\sqrt{1-\beta_\tau^2}z\sqrt{1-z^2}\\
	\end{cases}\, ,
	\label{eq:Coeff80}
\end{equation}

\begin{align}
	F^{[8,0]} &= 32 N e^2 \left( \frac{m_\tau v}{\Lambda} \right)^2 \frac{1}{1-\beta_\tau^2}\biggr( Q_f^2\Re(c_{\gamma})^2 - 8 \Re(c_{\gamma})\Re(c_{Z})\Re\left( \frac{Q_{f}Q_{Vf}m^2_\tau}{D_Z} \right) \\
		  &+16 \Re(c_{\gamma}^2)\frac{m_\tau^4 c_W^2s_W^2 (Q_{Af}^2 + Q_{Vf}^2)}{|D_Z|^2} \biggr) \nonumber \, ,\\
	F^{[8,0]}_B &= 64 N e^2 \left( \frac{m_\tau v}{\Lambda} \right)^2 \left( \Re(c_{\gamma})\Re(c_{Z})\Re\left( \frac{Q_{f}Q_{Af}m^2_\tau}{D_Z}\right) + 4\Re(c_{Z})^2\frac{m_\tau^4 c_W^2s_W^2 Q_{Af}Q_{Vf}}{|D_Z|^2} \right) \, .
\end{align}

The imaginary squared contribution is particularly simple; indeed it produces a singlet state
$C^{[8,2]} = - A^{[8,2]}\mathds{1}_3$, with the overall normalisation being
\begin{equation}
	A^{[8,2]}= F^{[8,2]} =  F^{[8,0]}\beta_\tau^2 (1-z^2) \; \textrm{with} \; \left( \Re(c_{\gamma/Z}) \to \Im(c_{\gamma/Z}) \right)\, .\\
\end{equation}

The pure real and pure imaginary contributions produce then two orthogonal spin configuration, which cannot interfere in the cross section:
\begin{equation}
	A^{[8,1]} = 0 \, .
	\label{eq:A81}
\end{equation}

Nonetheless, spin correlations resurrect the interference, and we obtain
\begin{equation}
	\begin{cases}
		\tilde{B}^{[8,1]}_r = F^{[8,1]}_{\Gamma,B}\dfrac{z\sqrt{1-z^2}}{\sqrt{1-\beta_\tau^2}} \\
		\tilde{B}^{[8,1]}_n = F^{[8,1]}\dfrac{\sqrt{1-z^2}}{\sqrt{1-\beta_\tau^2}}\\
		\tilde{B}^{[8,1]}_k = F^{[8,1]}_{\Gamma,B}1-z^2 \\
		\tilde{C}^{[8,1]}_{rn} = F^{[8,1]}\dfrac{1-z^2}{1-\beta_\tau^2} \\
		\tilde{C}^{[8,1]}_{rk} = F^{[8,1]}_\Gamma\dfrac{\sqrt{1-z^2}}{\sqrt{1-\beta_\tau^2}} \\
		\tilde{C}^{[8,1]}_{nk} = F^{[8,1]}\sqrt{1-\beta_\tau^2}z\sqrt{1-z^2} \\
	\end{cases} \, ,
	\label{eq:Coeff81}
\end{equation}
\begin{align}
	F^{[8,1]} &= 64 N e^2 \left( \frac{m_\tau v}{\Lambda} \right)^2 \frac{\beta_\tau}{1-\beta_\tau^2} \biggr( Q_f^2\Re(c_\gamma)\Im(c_\gamma) \\
		  & + 4 Q_fQ_{Vf} \left(\Re(c_{\gamma})\Im(c_{Z})+ \Re(c_{Z})\Im(c_{\gamma})\Re(\frac{c_W s_W m_\tau^2}{D_Z})\right) + 16\frac{c_W^4 s_W^4 (Q_{Vf}^2 + Q_{Af}^2)\Re(c_{Z})\Im(c_{Z})m_{\tau}^4}{|D_Z|^2}\biggr) \nonumber \, , \\
	F^{[8,1]}_\Gamma &= 128 N e^2 \left( \frac{m_\tau v}{\Lambda} \right)^2 \beta_\tau \left(Q_{f}Q_{Af}\Re(c_{\gamma})\Im(c_{Z}) - \Re(c_{Z})\Im(c_{\gamma})\Im(\frac{c_W^3s_W^3m_\tau^2}{D_Z})\right)\\
	F^{[8,1]}_B &= 256 N e^2 \left( \frac{m_\tau v}{\Lambda} \right)^2 \beta_\tau\biggr( Q_{f}Q_{Af}\left(\Re(c_{Z})\Im(c_{\gamma}) - \Re(c_{\gamma})\Im(c_{Z})\right)\Re(\frac{c_W s_W m_\tau^2}{D_Z}) \\
\nonumber &+ 4\frac{\Re(c_{Z})\Im(c_w^4 s_w^4 c_{Z})Q_{Af}Q_{Vf}m_\tau^4}{|D_Z|^2}\biggr)\, ,\\
F^{[8,1]}_{\Gamma,B} &= F^{[8,1]}_{\Gamma} \; \textrm{with} \; \left( Q_{Af} \to Q_{Vf}  \right)\, .
\end{align}

\section{Monte Carlo Simulations}
\label{sec:DetailsMC}

In this appendix, we present the results of the MC simulations for lepton colliders. The simulations are performed using MadGraph5\_aMC@NLO~\cite{Alwall:2014hca}, Pythia 8.3~\cite{Bierlich:2022pfr} and Delphes~\cite{deFavereau:2013fsa}, as well as the TauDecay package~\cite{Hagiwara:2012vz}. For all benchmark energies, we restrict to events where both taus decay through the same channel, either into one charged pion or one charged pion and one neutral pion. For the three-body case, the decay product $d$ in $\cref{eq:Tomography}$ is the $\rho/\rho'$ resonance, whose 4-momentum is the sum of the 4-momenta of the observable pions. We use the results in appendix A of Ref.~\cite{Bernreuther:2021elu} for the numerical values of the spin-analysing power:
\begin{equation}
	\begin{cases}
		\alpha_{\pi\nu} &= 1\\
		\alpha_{\pi\pi^0\nu} &=0.42
	\end{cases}\, ,
	\label{<+label+>}
\end{equation}
where we note that the computation of $\alpha_{\pi\pi^0\nu}$ takes into account the finite $\rho(\rho')$ width~\cite{Bernreuther:2021elu}. 

We generate 1 million events for each energy $\sqrt s = (m_\Upsilon,m_Z,500)$, and obtain statistical uncertainties by a bootstrap method~\cite{ATLAS:2021kho} with 30 pseudo-experiments, for both the Fano coefficients and the quantum information observables $\mathfrak{m}_{12}$ and $\mathcal{C}[\rho]$.
\subsection{Z-Pole}
Let us focus on the $Z$-pole first. We use the standard MadGraph parameters: 
\begin{equation}
	\begin{cases}
	m_\tau = 1.777\; \textrm{GeV}\\
	m_Z = 91.188\; \textrm{GeV}\\
	\Gamma_Z = 2.4414\; \textrm{GeV}\\
	s_W^2 = 0.222246\\
	\end{cases}
	\label{}
\end{equation}

and obtain, using the Fano coefficients in \cref{sec:Fano}:
\begin{align}
	B_+ = B_- &=
	\begin{pmatrix}
		0.003 & 0 & 0.2183
	\end{pmatrix} \\
C &=
	  \begin{pmatrix}
		  0.4825 & 0.008 & 0\\
		  0.008 & -0.4825 & 0.0011\\
		  0 & 0.0011 & 1
	  \end{pmatrix}\\
\mathfrak{m}_{12} &= 1.233 \\
\mathcal{C}[\rho] &= 0.471 \, .
	\label{<+label+>}
\end{align}

As a baseline, we perform simulations without ISR or detector effects. For the two-body decay channel, we find
\begin{align}
	B_- &=
	\begin{pmatrix}
	0.0009\pm0.0016 & 0.003\pm0.0019&0.218\pm0.002
	\end{pmatrix}\\
	B_+ &=
	\begin{pmatrix}
	0.0012\pm0.0016&0.004\pm0.0016&0.217\pm0.0018
	\end{pmatrix}\\
	C &=
	\begin{pmatrix}
	0.479\pm0.003& 0.012\pm0.003&-0.001\pm0.003\\
	0.008\pm0.003&-0.479\pm0.003&0.004\pm0.003\\
	0.001\pm0.003&0.0017\pm0.002&0.997\pm0.003\\
	\end{pmatrix}\\
	\mathfrak{m}_{12} &=1.227\pm0.007\\
	\mathcal{C}[\rho] &=0.476\pm0.003
\end{align}

The Fano coefficients are mostly in agreement within the quoted error with the analytical results, but some of the Fano coefficient are shifted downwards by up to 1 standard deviations. It should be noted that a similar shift is present in Ref.~\cite{Fabbrichesi:2024wcd}, but the uncertainty there is not small enough to make it evident. As a consequence, the concurrence is significantly shifted upwards, while the Horodecki marker agrees within the quoted uncertainty, while also shifted downwards with respect to the analytic computation.

For the three-body decay, we instead find
\begin{align}
B_- &=
\begin{pmatrix}
0.003\pm0.004&0.0019\pm0.004&0.205\pm0.005
\end{pmatrix}\\
B_+ &=
\begin{pmatrix}
0.0004\pm0.004&0.004\pm0.003&0.210\pm0.004
\end{pmatrix}\\
C &=
\begin{pmatrix}
0.451\pm0.017&-0.024\pm0.014&-0.041\pm0.017\\
-0.009\pm0.013& -0.49\pm0.02&0.015\pm0.013\\
-0.023\pm0.018&-0.015\pm0.015&0.98\pm0.02\\
\end{pmatrix}\\
\mathfrak{m}_{12} &=1.21\pm0.04\\
\mathcal{C}[\rho] &=0.444\pm0.016\;.
\label{eq:ThreeBodyTruth}
\end{align}
Even at truth-level, we find worse agreement from the analytical result with respect to the two-body analysis. Moreover, it is interesting to observe that, with the same number of events, the statistical uncertainty on the Fano coefficients is worse than the two-body case by a factor of roughly $6$. An explanation of this fact is beyond the scope of this paper. We note that the three-body decay has a higher braching ratio compared to the two-body: $BR(\tau\to\rho(\rho')\nu)/ BR(\tau\to\pi\nu) \approx 2.5$, but that does not compensate the worse intrinsic statistical uncertainty. In the following, and to derive the constraints in \cref{fig:plLambdas} we consider the two-body decay only.

We now introduce ISR effects, which are natively incorporated in Madgraph using the language of lepton PDFs~\cite{Frixione:2021zdp}. We obtain
the ISR changes the available energy in the hard collision from the nominal one, which breaks the assumption that all events have the same energy, introducing a systematic effect, see \cref{fig:MomDistr}.
The results in \cref{eq:TwoBodyTruth,eq:ThreeBodyTruth} are derived using truth-level quantities, and in particular the tau momenta are known exactly.
However, since the neutrino momentum cannot be measured at colliders, the reconstruction of the tau momentum has to be performed following certain assumptions.
We use the prescription of~\cite{Jeans:2015vaa,Altakach:2022ywa}, that resolves the two-fold ambiguity due to the nonobservance of the neutrino, but forces us to assume the value of the di-tau invariant mass, see \cref{sec:ImpactPar}.  
As a result, ISR introduces an unavoidable systematic effect. We can check the magnitude of this effect by integrating over the momentum distribution (see \cref{fig:MomDistr}). Using the analytic Fano Coefficients we obtain

\begin{align}
	B_+ = B_- &=
	\begin{pmatrix}
		0.003 & 0 & 0.2166
	\end{pmatrix} \\
	C&=
	  \begin{pmatrix}
		  0.4821 & 0.008 & 0\\
		  0.008& -0.4821 & 0.0007\\
		  0 & 0.0007 & 1
	  \end{pmatrix}\\
\mathfrak{m}_{12} &= 1.233 \\
\mathcal{C}[\rho] &= 0.471 \, .
	\label{eq:FanoISR}
\end{align}
the overall effect is rather modest, due to the (normalised) Fano coefficients being ratios of cross sections which are mostly affected the same way by ISR\@.

To also check the impact of detector effects, we perform a fast detector simulation using Delphes~\cite{deFavereau:2013fsa}, with the IDEA card developed for FCC-ee studies.
The results are
\begin{align}
B_- &=
\begin{pmatrix}
0.0025\pm0.0018&-0.008\pm0.002&0.2518\pm0.0012
\end{pmatrix}\\
B_+ &=
\begin{pmatrix}
0.0001\pm0.0017&-0.003\pm0.0015&0.252\pm0.0018
\end{pmatrix}\\
C &=
\begin{pmatrix}
0.504\pm0.003&0.008\pm0.003&-0.005\pm0.003\\
0.011\pm0.003&-0.477\pm0.003&0.0004\pm0.003\\
0.007\pm0.003&-0.007\pm0.003&1.03\pm0.003\\
\end{pmatrix}\\
\mathfrak{m}_{12} &=1.324\pm0.006\\
\mathcal{C}[\rho] &=0.476\pm0.003
\label{eq:TwoBodyTruth}
\end{align}
which are indeed shifted from the truth-level values in \cref{eq:TwoBodyTruth}, as well as the analytical results including ISR in~\cref{eq:FanoISR}. We note that the statistical uncertainties remain unchanced. We thus believe the statistical uncertainties derived here to be a good approximation of the statistical uncertainties that will arise in a realistic setting. We leave the question of the estimate of systematic uncertainties, which should be dominated by detector effects, to the experimental collaboration.

In what follows, we list the results for the benchmark energies $\sqrt{s} = 10.58\; \textrm{GeV}$ and $\sqrt s = 500\; \textrm{GeV}$. Given that we are mostly interested in the statistical uncertainties, we refrain from performing a detector simulation in these cases. Indeed, we are mostly interested in the statistical uncertainty, and for the reason explained above we believe that the truth-level MC simulation represents an accurate estimate.

\subsection{Belle II}
Belle II operates with asymmetric beams. However, it is always possible to Lorentz boost to the c.o.m.~frame, so we report the results in this frame. The analytic results are
\begin{align}
	B_+ = B_- &=
	\begin{pmatrix}
		0 & 0.0004 & -0.0016
	\end{pmatrix} \\
C &=
	  \begin{pmatrix}
		  -0.4199 & 0 & 0\\
		  0& 0.5267 & -0.0017\\
		  0 & -0.0017 & 0.8932 
	  \end{pmatrix}\\
\mathfrak{m}_{12} &= 1.075 \\
\mathcal{C}[\rho] &= 0.4451 \, .
	\label{}
\end{align}
at truth level, without ISR, we obtain:
\begin{align}
B_- &=
\begin{pmatrix}
0.0037\pm0.0014&0.000\pm0.0014&0.000\pm0.0016
\end{pmatrix}\\
B_+ &=
\begin{pmatrix}
0.0026\pm0.0019&0.000\pm0.0018&0.0016\pm0.0015
\end{pmatrix}\\
C &=
\begin{pmatrix}
-0.394\pm0.003&0.000\pm0.003&0.007\pm0.003\\
0.006\pm0.003&0.485\pm0.003&0.159\pm0.002\\
0.000\pm0.004&0.260\pm0.002&0.800632\pm0.003\\
\end{pmatrix}\\
\mathfrak{m}_{12} &=0.970\pm0.005\\
\mathcal{C}[\rho] &=0.373\pm0.003 \, ,
	\label{<+label+>}
\end{align}
which has worse agreement with the analytical result than the $\sqrt{s} = m_Z$ case.
\subsection{Linear Collider}

The analytic results are
\begin{align}
	B_+ = B_- &=
	\begin{pmatrix}
		0 & 0.0006 & 0.099
	\end{pmatrix} \\
C &=
	  \begin{pmatrix}
		  -0.3771 & 0 & 0\\
		  0& 0.3772 & 0.0027\\
		  0 & 0.0027 &  1
	  \end{pmatrix}\\
\mathfrak{m}_{12} &= 1.142 \\
\mathcal{C}[\rho] &= 0.375 \, .
	\label{}
\end{align}
at truth level, without ISR, we obtain:
\begin{align}
	B_- &=
\begin{pmatrix}
0.0030\pm0.0018&0.0000\pm0.0012&0.098\pm0.0015
\end{pmatrix}\\
B_+ &=
\begin{pmatrix}
0.000\pm0.0019&-0.0024\pm0.0017&0.099\pm0.0014
\end{pmatrix}\\
C &=
\begin{pmatrix}
-0.377\pm0.003&0.000\pm0.002& 0.000\pm0.003\\
0.000\pm0.003&0.375\pm0.003&0.005\pm0.004\\
0.000\pm0.003&0.004\pm0.003&0.998\pm0.003\\
\end{pmatrix}\\
\mathfrak{m}_{12} &=1.141\pm0.005\\
\mathcal{C}[\rho] &=0.373\pm0.002 \, .
	\label{fig:MomDistr500}
\end{align}
It should be noted that ISR effects are expected to be relevant here, because of the large energy of the beam. Any realistic prediction for the Fano coefficients should take this into consideration.
\section{Reconstruction of Tau Momenta}
\label{sec:ImpactPar}
Here, we collect for completeness the relevant formula from~\cite{Jeans:2015vaa,Altakach:2022ywa} to reconstruct the tau rest frame. 
At lepton colliders, in the absence of ISR, the momentum of the taus are constrained to sum to the center of mass energy. We use a not orthonormal basis to express them: ${p_{\textit{c.o.m.}}^\mu,p_{\pi^+}^\mu,p_{\pi^-}^\mu,q^\mu}$, where $p_{\textit{c.o.m.}} = (E,0,0,0)$ and
	\begin{equation}
		q^{\mu} = \frac{1}{s}\epsilon^{\mu\nu\rho\sigma}p_{\textit{c.o.m.}}^{\nu}p_{\pi^+}^\rho p_{\pi^-}^\sigma	
	\label{<+label+>}
\end{equation}
The tau momenta can then be expressed in this basis as:
\begin{equation}
	p_{\tau^\pm}^\mu = \frac{1\mp a}{2}p_{\textit{c.o.m.}}^\mu \pm \frac{b}{2} p_{\pi^+}^\mu \mp \frac{c}{2}p_{\pi^-}^\mu\pm d q^\mu
	\label{<+label+>}
\end{equation}
The coefficients can be obtained by imposing that all the particles are on-shell:
\begin{equation}
	\begin{cases}
		p_{\tau^+} = m^2_{\tau}\\
		p_{\tau^-} = m^2_{\tau}\\
		(p_{\tau^+}^\mu - p_{\pi^+})^2 = m_\nu^2 \approx 0\\
		(p_{\tau^-}^\mu - p_{\pi^-})^2 = m_{\bar\nu}^2 \approx 0\\
	\end{cases}\, .
	\label{<+label+>}
\end{equation}
Carrying out the algebra is easy to see that $a,b,c$ can be obtained as solutions of the following matrix equation~\cite{Altakach:2022ywa}:
\begin{equation}
	\begin{pmatrix}
		-x& m_{\pi}^2&z\\
		y& -z & m_{\pi}^2\\
		s& -x& y
	\end{pmatrix}
	\begin{pmatrix}
		a\\
		b\\
		c\\
	\end{pmatrix}
	=
	\begin{pmatrix}
		m_{\tau}^2 + m_{\pi}^2 - x\\
		m_{\tau}^2 + m_{\pi}^2 - y\\
		0
	\end{pmatrix}
	\label{<+label+>}
\end{equation}
where 
\begin{equation}
	\begin{cases}
		x = p_{\textit{c.o.m.}}\cdot p_{\pi^+}\\
		y = p_{\textit{c.o.m.}}\cdot p_{\pi^-}\\
		z = p_{\pi^+}\cdot p_{\pi^-}\\
	\end{cases}
	\label{<+label+>}
\end{equation}
and, finally, $d$ is the solution to
\begin{equation}
	d^2 = -\frac{1}{4q^2}\left( \left( 1+a \right)s + (b^2+c^2)m_{\pi}^2 - 4m_{\tau}^2 + 2\left( acy - abx - bcz \right) \right) \,.
	\label{<+label+>}
\end{equation}
There are two possible solution to this equation, an ambiguity resulting from the nonobservance of the neutrinos. It is also clear that, if $p_{\textit{c.o.m.}}^\mu$ is not known, this procedure cannot be carried out, as is the case when the initial state is composite. 

The ambiguity can be partially resolved by using information from the impact parameter $\vec{b}_{\pm}$, that is the minimal displacement of the $\pi^{\pm}$ flight direction from the interaction point, which we set to be the origin. This can be measured experimentally, $\vec{b}_{\textit{meas}}$, or calculated from the reconstructed tau momentum and measured pi momentum, $\vec{b}_{\textit{reco}}$. Given the ambiguity on $d$, there are two values of $\vec{b}_{\textit{reco}}^{\pm}$. We choose the solution of $d$ which minimises
\begin{equation}
	\Delta = \left|\vec{b}_{\textit{meas}}^{+}-\vec{b}_{\textit{reco}}^{+}\right| + \left|\vec{b}_{\textit{meas}}^{-}-\vec{b}_{\textit{reco}}^{-}\right| \, . 
	\label{<+label+>}
\end{equation}
In~\cite{Altakach:2022ywa} a likelihood method that takes into account the experimental uncertainties as nuisance parameters is used, but we find that simply picking the solution with the smallest $\Delta$ works for the purposes of this paper. 

Finally, although this algorithm is developed for two-body decays, the extension to $\tau \to (\rho/\rho')\nu \to \pi\pi^0 \nu$ is trivial: one simply substitutes $p_{\pi^\pm}^\mu$ for $p_{\rho^{\pm}}^\mu \equiv p_{\pi^\pm}^\mu + p_{\pi^0}^\mu$ everywhere.
\bibliographystyle{apsrev4-1}
\bibliography{Refs}
\end{document}